%% file: ms.tex
\documentclass{vldb}

\newcommand {\xpointshort} {Optane PMM }
\newcommand {\xpointfull} {Intel\textsuperscript{\textregistered} Optane\textsuperscript{TM} DC Persistent Memory }
\newcommand {\xpointshortend} {Optane PMM}

\usepackage{url,afterpage,amssymb,amsfonts}
\usepackage{enumitem,makecell}
\usepackage{booktabs}
\usepackage{times}
\usepackage{multirow}
\usepackage{subfig}
\usepackage{graphicx}
\usepackage{color}
\usepackage{colortbl}
\usepackage{balance}
\usepackage{listings}
\usepackage{xcolor}
\usepackage{colortbl}
\usepackage{hyphenat}

\newcommand{\revise}[1]{{\color{black} {#1}}}

\newenvironment{closeitemize}{\begin{list}{\small $\bullet$}%
{\usecounter{enumiv} \setlength{\itemsep}{0in} \setlength{\parsep}{0in}
\setlength{\topsep}{1ex}}
\def\makelabel##1{\hss\llap{##1}}}%
{\end{list}}

\def\BibTeX{{\rm B\kern-.05em{\sc i\kern-.025em b}\kern-.08emT\kern-.1667em\lower.7ex\hbox{E}\kern-.125emX}}

\graphicspath{{figs/}}

\vldbTitle{Single Machine Graph Analytics on Massive Datasets Using Intel Optane Persistent Memory}
\vldbAuthors{Gurbinder Gill, Roshan Dathathri, Loc Hoang, Ramesh Peri, and Keshav Pingali}
\vldbDOI{https://doi.org/10.14778/xxxxxxx.xxxxxxx}
\vldbVolume{12}
\vldbNumber{xxx}
\vldbYear{2019}

\begin{document}

\title{Single Machine Graph Analytics on Massive Datasets Using Intel Optane DC Persistent Memory}

\author{
%
%
\alignauthor
Gurbinder Gill$^1$, Roshan Dathathri$^1$, Loc Hoang$^1$, Ramesh Peri$^2$, Keshav Pingali$^1$\\
\affaddr{University of Texas at Austin$^1$},
\affaddr{Intel Corporation$^2$}
\email{\normalsize\{gill,roshan,loc,pingali\}@cs.utexas.edu,
\normalsize\{ramesh.v.peri\}@intel.com}
}

\date{}

\maketitle

\input{abstract}

\section{Introduction}
\label{sec:intro}
\input{intro}


\section{\xpointshort}
\label{3dxpoint}
\input{3dxpoint}

\section{Platforms and Graph Analytics Systems}
\label{subsec:experimental_setup}
\input{results_setup}

\section{Memory Hierarchy Issues}
\label{sec:mem-access}
\label{subsec:3dxpoint_setup}
\input{mem_access}


\section{Efficient Algorithms for Massive Graphs}
\label{subsec:algo_design}
\label{sec:abstractions}
\input{operator}

\section{Evaluation of Graph Frameworks}
\label{sec:results}
\input{results}

\section{Related Work}
\label{sec:related}
\input{related}

\section{Conclusions}
\label{sec:conclusions}
\input{conclusion}

\bibliographystyle{abbrv}
\bibliography{related,iss,numa,others,graphs,gpugraphs,partitioning,fpgagraphs,nvm,outofcore,dimitri,tools}

\end{document}

%% file: abstract.tex
\begin{abstract}

Intel Optane DC Persistent Memory (Optane PMM) is a new kind of
byte-addressable memory with higher density and lower cost than DRAM. This
enables the design of affordable systems that support up to 6TB of randomly
accessible memory. In this paper, we present key runtime and algorithmic
principles to consider when performing graph analytics on extreme-scale graphs
on Optane PMM and highlight principles that can apply to graph analytics
on all large-memory platforms.

To demonstrate the importance of these principles, we evaluate four existing
shared-memory graph frameworks and one out-of-core graph framework on large
real-world graphs using a machine with 6TB of Optane PMM. Our results show
that frameworks using the runtime and algorithmic principles advocated in
this paper (i) perform significantly better than the others and (ii) are
competitive with graph analytics frameworks running on large production
clusters. 

\end{abstract}

%% file: intro.tex
Graph analytics systems must process graphs with tens of billions of nodes
and trillions of edges. Since the main memory of most single machines is
limited to a few hundred GBs, shared-memory graph analytics systems like
Ligra~\cite{ligra}, Galois~\cite{galois}, and GraphIt~\cite{graphit} cannot be
used to perform in-memory processing of these large graphs. Two approaches
have been used in the literature to circumvent this problem: (i)
\emph{out-of-core} processing and (ii) \emph{distributed-memory} processing.

In out-of-core systems, the graph is stored in secondary storage (SSD/disk),
and portions of the graph are read into DRAM under software control for
in-memory processing. State-of-the-art systems in this space include
X-Stream~\cite{xstream}, GridGraph~\cite{gridgraph}, Mosaic~\cite{mosaic}, and
BigSparse~\cite{bigsparse}. Secondary storage devices do not support random
accesses efficiently: data must be fetched and written in blocks. As a
consequence, algorithms that perform well on shared-memory machines often
perform poorly in an out-of-core setting, and it is necessary to rethink
algorithms and implementations when transitioning from in-memory graph
processing to out-of-core processing. In addition, the graph may need to be
preprocessed to organize the data into an out-of-core friendly layout.

Large graphs can also be processed with distributed-memory clusters.
The graph is partitioned among the machines in a cluster using one of many
partitioning policies in the literature~\cite{partitioningstudy}. Communication
is required during the computation to synchronize node updates.
State-of-the-art systems in this space include D-Galois~\cite{gluon} and
Gemini~\cite{gemini}. Distributed-memory graph analytics systems have the
advantage of scale out by adding new machines to provide
additional memory and compute power. The overhead of
communication can be reduced by choosing good partitioning policies, avoiding
small messages, and optimizing metadata, but communication remains the
bottleneck in these systems~\cite{gluon}. Obtaining access to large
clusters may also be too expensive for many users.

\xpointfull (\xpointshortend) is new byte-addressable memory technology
with the same form factor as DDR4 DRAM modules with higher memory density and
lower cost. It has longer access times compared to DRAM, but it is much faster
than SSD. 
It allows a single machine to have up to 6TB of storage at relatively low cost,
and in principle, it can run memory-hungry applications without requiring the
substantial reworking of algorithms and implementations needed by out-of-core
or distributed-memory processing.

We explore the use and viability of \xpointshort for analytics of very large
graphs such as web-crawls up to 1TB in size.  We design and present studies
conducted to determine \emph{how} to run graph analytics applications
on \xpointshort and large-memory systems in general. 
Our studies make the following points:

\begin{enumerate}
  \item \revise{Non-uniform memory access (NUMA)}-aware memory allocation of
  graph data structures that maximizes near-memory (DRAM treated as cache)
  usage is important on \xpointshort as cache misses on the platform are
  significantly slower than cache misses on DRAM.
  (Section~\ref{subsec:3dxpoint_setup})

  \item Avoiding page management overhead while using \xpointshort is
  key to performance as kernel overhead on \xpointshort is higher due to higher
  access latency. (Section~\ref{subsec:3dxpoint_setup})

  \item Algorithms must avoid high amounts of memory accesses on large memory
  systems: graph frameworks should give users the flexibility to write
  non-vertex, asynchronous programs with efficient parallel data structures.
  (Section~\ref{subsec:algo_design})
\end{enumerate}

We evaluate four shared-memory graph analytics frameworks --
Galois~\cite{galois}, GAP~\cite{gap}, GraphIt~\cite{graphit}, and
GBBS~\cite{gbbs} -- on \xpointshort to show the importance of these
practices for graph analytics on large-memory systems. We compare the
performance of the best of these frameworks, Galois, with the state-of-the-art
distributed graph analytics system, D-Galois, and show that a system running on
\xpointshort that uses our proposed practices is competitive with the
same algorithms run on D-Galois with 256 machines, and since the
\xpointshort system supports more efficient shared-memory algorithms
which are difficult to implement on distributed-memory machines, applications using the
more efficient algorithms can outperform distributed-memory execution.  We
also evaluate the out-of-core graph analytics system GridGraph~\cite{gridgraph}
using \xpointshort as external storage in app-direct mode (explained in
Section~\ref{3dxpoint}) and show that using \xpointshort as main memory in
memory mode (explained in Section~\ref{3dxpoint}) is orders of magnitude faster
than app-direct as it
allows for more sophisticated algorithms from shared-memory graph analytics
systems that out-of-core systems currently do not support (in particular,
non-vertex programs and asynchronous data-driven algorithms \revise{detailed in
Section~\ref{sec:abstractions}}).



The paper is organized as follows. Section~\ref{3dxpoint}
introduces \xpointshortend. Section~\ref{subsec:experimental_setup} describes
the experimental setup used in our studies.
Section~\ref{sec:mem-access} describes how to efficiently use the memory
hierarchy on large-memory systems and \xpointshort for graph analytics.
Section~\ref{sec:abstractions} discusses graph algorithm design for
large-memory systems. Section~\ref{sec:results} presents our
evaluation. Section~\ref{sec:related} surveys related work. \\

%% file: 3dxpoint.tex
\begin{figure}[t]
    \centering
    \includegraphics[width=0.48\textwidth]{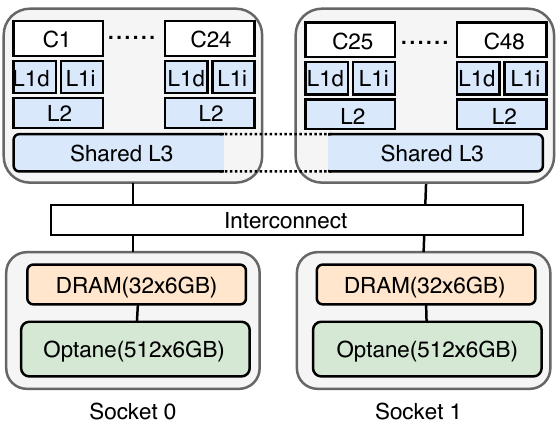}
    \caption{Memory hierarchy of our 2 socket machine with 384GB of DRAM and 6TB of Intel Optane PMM.}
  \label{fig:3dxpoint_diagram}
   \vspace{-10pt}
\end{figure}

\begin{figure}
  \centering
  \includegraphics[width=0.48\textwidth]{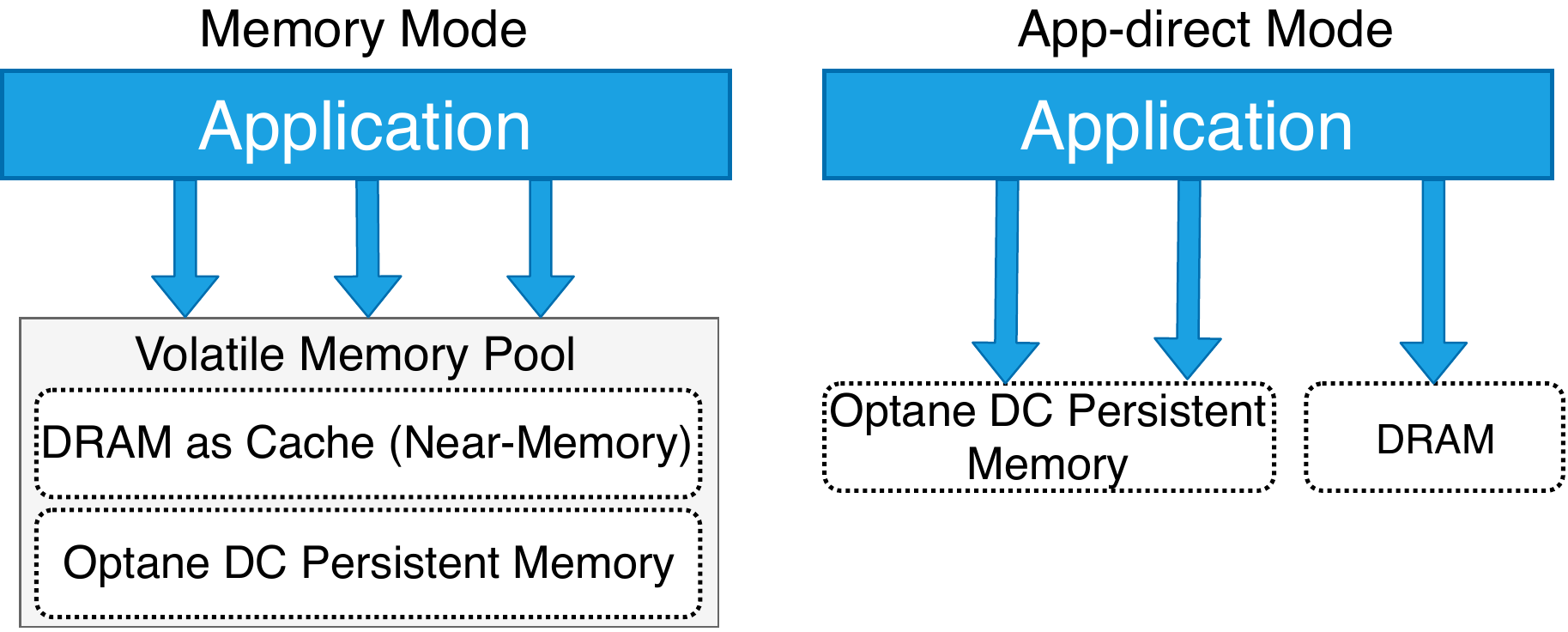}
  \caption{Modes in \xpointshortend.}
\label{fig:3dxpoint_modes}
\end{figure}

\xpointshort delivers a combination of
affordable large capacity and persistence (non-volatility). As shown in
Figure~\ref{fig:3dxpoint_diagram}, \xpointshort adds a new level to the
memory hierarchy. It comes in the same form factor as a DDR4 memory
module and has the same electrical and physical interfaces. However, it uses a
different protocol than DDR4 which means that the CPU must have \xpointshort
support in its memory controller. Similar to the DRAM distribution in
\revise{non-uniform memory systems in which memory is divided into sockets},
the \xpointshort modules are distributed among sockets.
Figure~\ref{fig:3dxpoint_diagram} shows an example of a two socket machine with
6TB of \xpointshort split between sockets.
%
%
\xpointshort can be configured as volatile main memory (memory mode),
persistent memory (app-direct mode), or a combination of both
(Figure~\ref{fig:3dxpoint_modes}).

\textbf{Memory Mode}: In memory mode, \xpointshort is treated as
as main memory, and \revise{DRAM acts as direct-mapped (physically indexed and
physically tagged) cache called \emph{near\hyp{}memory}}.  The granularity of
caching from \xpointshort to DRAM is 4KB.
This enables the system to deliver DRAM-like performance at substantially lower
cost and power \revise{with no modifications to the application}. Although the
memory media is persistent, the software sees it as volatile memory. This
enables
systems to provide up to 6TB of randomly accessible storage, which is 
expensive to do with DRAM. 

Traditional code optimization techniques 
can be used to tune applications to run well in this configuration. In
addition, software must consider certain asymmetries in machines
with \xpointshortend. \xpointshort modules on a socket can use only the
DRAM present in its local NUMA node \revise{(i.e., socket)} as
near\hyp{}memory. Therefore, in addition to NUMA allocation considerations,
software using \xpointshort has to account for near\hyp{}memory hit rate 
as the cost of a local near\hyp{}memory miss is
higher than the remote near\hyp{}memory hit (discussed in
Section~\ref{subsec:3dxpoint_setup}).  Therefore, it should allocate
memory so that the system can utilize more DRAM as near-memory even if it means
more remote NUMA accesses.

\textbf{App-direct Mode}: In app-direct mode, \xpointshort modules are
treated as byte-addressable persistent memory. 
One compelling case for app-direct mode is in large memory databases where
indices can be stored in persistent memory to avoid rebuilding them on reboot,
achieving a significant reduction in restart time.
\xpointshort modules can be managed using an
API or a command line interface provided by the
\texttt{ipmctl}~\cite{ipmctl} OS utility in Linux.
\texttt{ipmctl}
can be used to 
configure the machine to use \emph{x}\% of \xpointshort modules capacity in the
memory mode and the rest in the app-direct mode
\footnote{\texttt{ipmctl create -goal MemoryMode=x PersistentMemoryType=AppD\hyp{}irect}};
for \emph{x} $>$ 0, all DRAM
on the machine is used as the cache (\texttt{near\hyp{}memory}). When all
the \xpointshort modules are in app-direct mode,
DRAM is the main volatile memory.

Specifications for the \xpointshort machine used in
our study are in Section~\ref{subsec:experimental_setup}.
Tables~\ref{tbl:bandwidth} and~\ref{tbl:latency} show the bandwidth and latency
of PMM observed on our machine.  Although \xpointshort is slower than
DDR4, its large capacity enables us to
analyze much larger datasets on a single machine than previously possible.  In
this paper, we focus on memory mode; we use app-direct mode for running
the out-of-core graph analytics system GridGraph~\cite{gridgraph}.

\input{tbl_optane_specs}
\input{tbl_optane_latency}

%% file: tbl_optane_specs.tex
\begin{table}[t]
\footnotesize
\centering

\begin{tabular}{c|c|rr|rr}

\toprule
\multirow{2}{*}{\textbf{Mode}}                & \multicolumn{1}{l|}{} & \multicolumn{2}{c|}{\textbf{Read}}                                           & \multicolumn{2}{c}{\textbf{Write}}                                          \\ \cline{3-6} 
               & \multicolumn{1}{l|}{} & \multicolumn{1}{c}{\textbf{Local}} & \multicolumn{1}{c|}{\textbf{Remote}} & \multicolumn{1}{c}{\textbf{Local}} & \multicolumn{1}{c}{\textbf{Remote}} \\
\midrule
\multirow{2}{*}{\textbf{Memory}}       & \textbf{Random}       & 90.0                               & 34.0                                    &  50.0                         &  29.5                             \\
                                    & \textbf{Sequential}   & 106.0                              & 100.0                                   &  54.0                         &  29.5                             \\ \hline
\multirow{2}{*}{\textbf{App-direct}} & \textbf{Random}       & 8.2                                & 5.5                                     &  3.6                           & 2.3                               \\
                                    & \textbf{Sequential}   &  31.0                         & 21.0                                    &  10.5                          & 7.5                               \\
\bottomrule
\end{tabular}
  \caption{Bandwidth (GB/s) of Intel Optane PMM.}
\label{tbl:bandwidth}
\vspace{-5pt}
\end{table}

%% file: tbl_optane_latency.tex
\begin{table}[t]
\footnotesize
\centering
\begin{tabular}{c|rr}
\toprule
\multicolumn{1}{c|}{\textbf{Mode}} & \multicolumn{1}{c}{\textbf{Local}} & \multicolumn{1}{c}{\textbf{Remote}} \\
\midrule
\multirow{1}{*}{\textbf{Memory}}         & 95.0                               & 150.0                             \\
                                    
\multirow{1}{*}{\textbf{App-direct}}    & 164.0                              & 232.0                              \\
                                    \bottomrule
\end{tabular}
\caption{Latency (ns) of Intel Optane PMM.}
\label{tbl:latency}
 \vspace{-5pt}
\end{table}

%% file: results_setup.tex
\xpointshort experiments were conducted on a 2 socket machine with Intel's
second generation Xeon scalable processor ("Cascade Lake") with 48 cores
(up to 96 threads with hyperthreading) with a clock rate of 2.2 Ghz. The
machine has 6TB of \xpointshortend, 384GB of DDR4 RAM, and 32KB L1, 1MB L2, and
33MB L3 data caches (Figure~\ref{fig:3dxpoint_diagram}).
The system has a 4-way associative data TLB with 64 entries for 2KB pages (small pages),
32 entries for 2MB pages (huge pages), and 4
entries for 1GB pages. Code is compiled with g++ 7.3. We used the same
machine for DRAM experiments by configuring the \xpointshort modules
to run entirely in app-direct mode and use DRAM as the
main volatile memory (equivalent to removing the \xpointshort
modules). We also use \xpointshort modules in app-direct mode to run GridGraph,
an out-of-core graph analytics system: DRAM is used as main memory and
\xpointshort modules are used as external storage.
{\em Transparent Huge Pages (THP)}\revise{~\cite{thp}, which tries to allocate huge 
pages for an application without explicitly reserving memory for huge pages}, are enabled
(default in Linux; 
).  To collect hardware counters and analyze performance, we
used Intel's Vtune Amplifier~\cite{vtune} and Platform Profiler~\cite{vpp}.

To show that our study of algorithms for massive graphs
(Section~\ref{subsec:algo_design}) is independent of machine architecture, we
also conducted experiments on a large DRAM 4 socket machine we call
Entropy. Entropy uses Intel Xeon Platinum 8176 ("Skylake") processors with a
total of 112 cores with a clock rate of 2.2 Ghz, 1.5TB of DDR4 DRAM, and 32KB
L1, 1MB L2, and 38MB L3 data caches. Code is compiled with g++ 5.4. For
our experiments on Entropy, we use 56 threads,
restricting our experiments to 2 sockets. 

\input{tbl_inputs}

Distributed-memory experiments were conducted on the Stampede2~\cite{stampede}
cluster at the Texas Advanced Computing Center using up to 256 Intel Xeon
Platinum 8160 ("Skylake") 2 socket machines with 48 cores with a clock rate of
2.1 Ghz, 192GB DDR4 RAM, and 32KB L1, 1MB L2, and 33MB L3 data caches. The
machines are connected with a 100Gb/s Intel Omni-Path interconnect.
Code is compiled with g++ 7.1.

Table~\ref{tbl:inputs} specifies the input graphs: clueweb12~\cite{clueweb},
uk14~\cite{BoVWFI,BRSLLP}, and wdc12~\cite{wdc} are web-crawls (wdc12 is the
largest publicly available one), and \revise{iso\_m100~\cite{protein} is a 
protein-similarity network 
(iso\_m100 is the largest dataset  
in IMG isolate genomes publicly available as part of the 
HipMCL software~\cite{protein}}). kron30 and rmat32
are randomized scale-free graphs generated using kron~\cite{kron} and
rmat~\cite{rmat} generators (using weights of 0.57, 0.19, 0.19, and 0.05, as
suggested by graph500~\cite{graph500}). \revise{Table~\ref{tbl:inputs} also
lists graph sizes on disk in Compressed Sparse Row (CSR) binary format}. kron30 and clueweb12 fit into DRAM, so
we use them to illustrate differences in workloads that fit into DRAM and those
that do not. The other graphs -- uk14, rmat32, and wdc12 -- do not fit in DRAM
on our \xpointshort machine.
We observe that
uk14 and wdc12 have non-trivial diameters, whereas rmat32 has a very small
diameter. We believe that rmat32 does not represent real-world datasets, so we
exclude it in all our experiments except to show the impact of diameter in our
study of algorithms (Section~\ref{subsec:algo_design}). All graphs are
unweighted, so we generate random weights.

Our evaluation uses 7 benchmarks: \revise{single-source betweenness centrality
(bc)~\cite{mrbc}, breadth-first search (bfs)~\cite{cormen01}, connected
components (cc)~\cite{lpcc,pjcc}, k-core decomposition (kcore)~\cite{kcore},
pagerank (pr)~\cite{pagerank}, single-source shortest path
(sssp)~\cite{meyer98}, and triangle counting (tc)~\cite{disttc}}.  The only
benchmark that uses \revise{edge} weights is sssp. The source node for bc, bfs, and sssp is
the maximum out-degree node. The tolerance for pr is $10^{-6}$. The $k$ in
kcore is $100$. \textit{All benchmarks are run until convergence except for pr,
which is run for up to 100 rounds}. We present the mean of 3 runs for the main
experiments.

The shared-memory graph analytics frameworks we use are Galois~\cite{galois},
GAP~\cite{gap}, GraphIt~\cite{graphit}, and GBBS~\cite{gbbs} (all described in
more detail in Section~\ref{sec:results}). D-Galois~\cite{gluon} is a
distributed-memory framework.  GridGraph~\cite{gridgraph} is an out-of-core
framework that streams graph topology and data into memory from external
storage (in this case, \xpointshort in app-direct mode).


%% file: tbl_inputs.tex
\begin{table}[t]
\footnotesize
\centering

\resizebox{0.48\textwidth}{!}{
\begin{tabular}{@{}l@{\hskip1.8pt}r@{\hskip 0.07in}r@{\hskip 0.0in }r@{\hskip 0.07in }r@{ }r@{ }r@{ }r@{}r@{ }r@{ }r@{ }}
\toprule
                               & \textbf{kron30}    & \textbf{clueweb12}       & \textbf{uk14}           & \textbf{iso\_m100}   & \textbf{rmat32}    & \textbf{wdc12}                   \\
\midrule                                                                                                                                             
\textsf{$|V|$}                 & 1,073M             & 978M                      & 788M            & 76M          & 4295M                    & 3,563M                           \\
\textsf{$|E|$}                 & 10,791M            & 42,574M                   & 47,615M         & 68,211M     & 68,719M                & 128,736M                         \\
\textsf{$|E|/|V|$}             & 16                 & 44                        & 60              & 896              & 16                   & 36                               \\
\textsf{max $D_{out}$}         & 3.2M               & 7,447                     & 16,365          &  16,107           & 10.4M             & 55,931                           \\
\textsf{max $D_{in}$}          & 3.2M               & 75M                       & 8.6M            & 31,687       & 10.4M                    & 95M                              \\
\textsf{Est. diameter}         & 6                   & 498                       & 2498           & 83            & 7                 &  5274                                 \\
\textsf{Size (GB)}             & 136                 &325                        & 361            & 509            &  544             &986                              \\
\bottomrule
\end{tabular}
}
\caption{Inputs and their key properties.}
\label{tbl:inputs}
\vspace{-5pt}
\end{table}

%% file: mem_access.tex
This section shows that on \xpointshort machines, the overhead of memory
operations such as \revise{non-uniform memory accesses (NUMA) across memory
sockets}, cache misses handling, and page table maintenance are higher than
DRAM machines. These overheads are reduced by intelligent memory allocation and
by reducing the time spent in the kernel for page-table maintenance. We address
three issues: NUMA-aware allocation (Section~\ref{alloc-numa-allocation}),
NUMA-aware migration (Section~\ref{alloc-numa-migration}), and page size
selection (Section~\ref{alloc-page-size}).

\begin{figure}
\footnotesize
\centering
\includegraphics[width=0.48\textwidth]{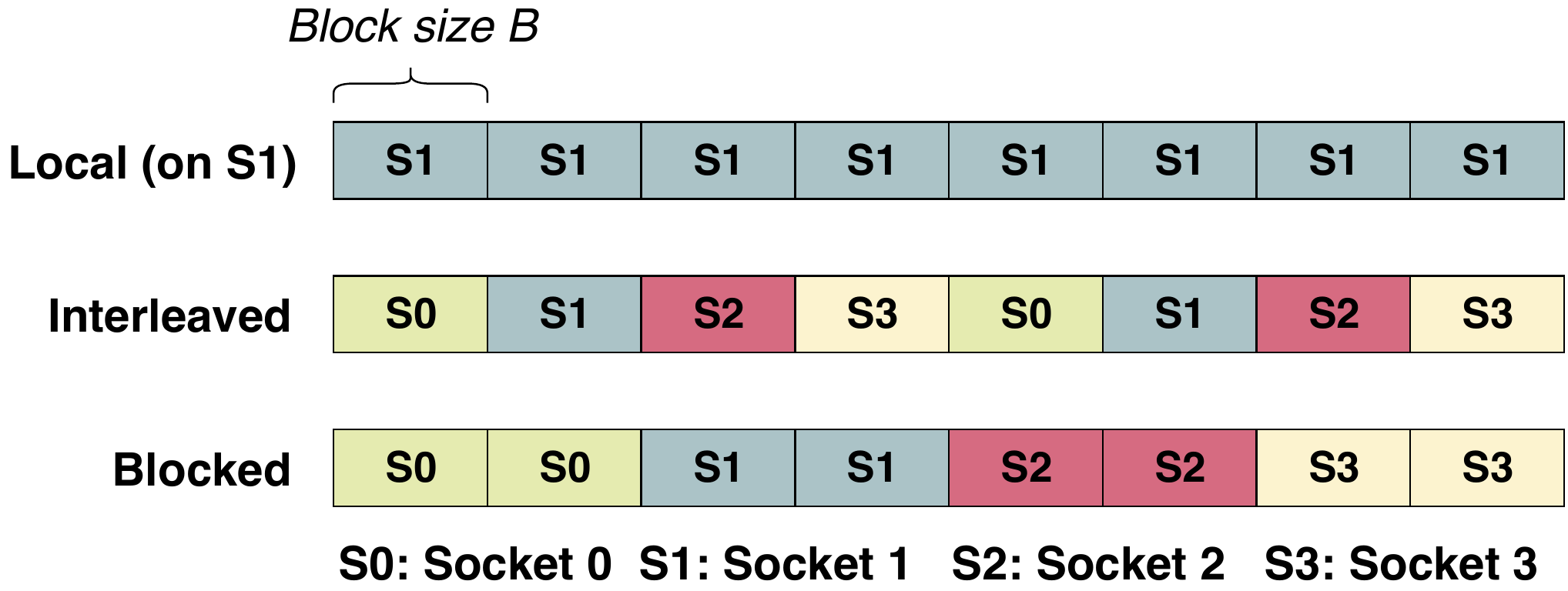}
\caption{Illustration of 3 different NUMA allocation policies on a 4-socket
system: each policy distributes blocks (size $B$, which is the size of a
page) of allocated memory among sockets differently.}
\label{fig:numa_policy}
\vspace{-10pt}
\end{figure}

\begin{figure}
\footnotesize
\centering
\begin{minipage}[t]{0.47\textwidth}
\begin{minipage}[t]{0.49\textwidth}
\centering
\includegraphics[width=\textwidth]{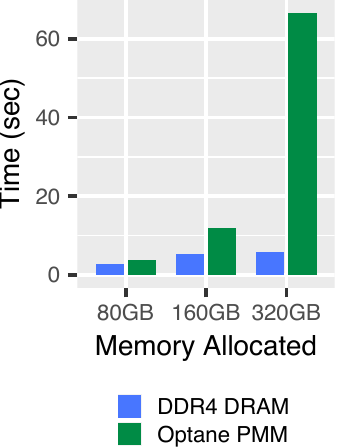}
(a)  NUMA local on 96 threads
\end{minipage}
\begin{minipage}[t]{0.49\textwidth}
\centering
\includegraphics[width=\textwidth]{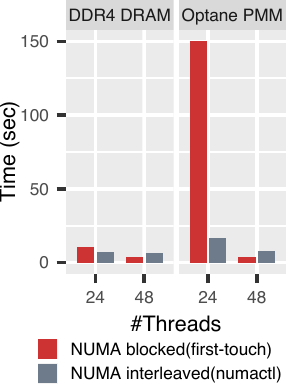}
(b) 320GB memory allocated
\end{minipage}
\end{minipage}
\caption{Time to write memory allocated on \xpointshort and DDR4 DRAM
using a micro-benchmark.}
\label{fig:res:numa_micro}
\vspace{-5pt}
\end{figure}

\subsection{NUMA-aware Allocation}
\label{alloc-numa-allocation}
\label{subsub:numa_policy}
%

NUMA-aware allocation increases bandwidth and reduces latency of memory accesses
by allocating memory on the same NUMA node as the cores that are likely to
access it.  Allocation falls into three main categories: (a)
\emph{NUMA local}, which allocates memory on a node specified at allocation
time (if there is not enough memory available on the preferred node, other
nodes are used), (b) \emph{NUMA interleaved}, which interleaves pages across
nodes in a round-robin fashion, and (c) \emph{NUMA blocked}, which blocks
the pages and distributes the blocks among nodes (illustrated in
Figure~\ref{fig:numa_policy}).

There are several ways for application programs to specify the allocation
policy. The policy can be set globally by using OS utilities such as
\texttt{numactl}\revise{~\cite{numactl}} on Linux. To allow different policies to be used in different
allocations, applications can use the OS-provided NUMA allocation library
(\texttt{numa.h} in Linux), which contains a variety of \texttt{numa\_alloc}
functions.  OS-based approaches, however, can only use the NUMA local or
interleaved policies. Another way to get fine-grained NUMA-aware allocation is
to manually allocate memory using \texttt{anonymous mmap}\revise{~\cite{mmap}} and have threads on
different sockets inside the application touch the pages (called
\texttt{first-touch}) to allocate them on the desired NUMA nodes. This method,
unlike OS-provided methods, allows applications to implement
application-specific NUMA-aware allocation policies.


To understand the differences in local, interleaved, and blocked NUMA
allocation policies on our \xpointshort setup, we use a micro-benchmark
that allocates different amounts of memory using different NUMA allocation
policies and writes to each location once using $t$ threads where each thread
gets a contiguous block to write sequentially. To explore the effects of NUMA
on different platforms, we run this microbenchmark with two setups: one using
only DDR4 DRAM (by setting \xpointshort into app-direct mode) and one using
\xpointshortend.
The following micro-benchmark results show that on the
\xpointshort machine, applications must not only maximize local NUMA accesses,
but \emph{must also use a NUMA policy that maximizes the near-memory
used in order to reduce DRAM conflict misses} (recall that DRAM is 
direct-mapped cache called near-memory for \xpointshort in this mode, \revise{and
limited cache size increases the probability of conflict misses due to physical addresses
mapping to the same cache line}).

\noindent{\bf NUMA Local.}
Figure~\ref{fig:res:numa_micro}(a) shows the execution time of the
microbenchmark on DDR4 DRAM and \xpointshort for the NUMA local allocation
policy using $t=96$ and different allocation amounts. Using NUMA local, all the
memory of socket 0 is used before memory from socket 1 is allocated. We observe
that going from 80GB to 160GB increases the execution time by $2\times$ for
both DRAM and \xpointshortend: this is expected since we increase the
work by $2\times$. Going from 160GB to 320GB also increases the work by
$2\times$. For DRAM, a 320GB allocation spills to the other socket (each socket
has only 192GB), increasing the effective bandwidth by $2\times$, so
the execution time does not change much.  In \xpointshort, however, the 320GB
is allocated entirely on socket 0 as our machine has 3TB per socket. Since
there is no change in bandwidth, one would expect the performance to degrade by
$2\times$, but it degrades by $5.6\times$. This is because the machine
can only use 192GB of DRAM as near-memory; this cannot fit 320GB, so the
conflict miss rate of the DRAM accesses increase by roughly 1.8$\times$. This
illustrates that (i) near-memory conflict misses are detrimental to the
performance for \xpointshort and (ii) NUMA local is not suitable for
allocations larger than 192GB on our setup.

\noindent{\bf NUMA Interleaved and Blocked.}
The execution times of the microbenchmark on DDR4 DRAM and \xpointshort for the
NUMA interleaved and blocked allocation policies using an allocation of 320GB
and different thread counts are shown in Figure~\ref{fig:res:numa_micro}(b). For
DRAM, both policies are similar for different $t$. When $t \leq 24$ on
\xpointshortend, NUMA blocked only allocates memory on socket 0 (because it
uses \texttt{first-touch}), so performance degrades 39$\times$ compared to 48
thread execution as 320GB does not fit in the near-memory of a single socket.
This illustrates that the cost of local near-memory misses is much higher than
the cost of remote near-memory hits.
In contrast, the NUMA interleaved policy for 24
threads uses both sockets and improves performance by 9$\times$ over NUMA blocked
even though 50\% of accesses are remote when $t \leq 24$.  NUMA interleaved
performs worse than blocked when $t=48$, as both allocation policies are able
to fit 320GB in the near-memory of 2 sockets (384GB); however, NUMA interleaved
results in more remote accesses as compared to NUMA blocked.


\subsection{NUMA-aware Migration}
\label{alloc-numa-migration}
\label{subsubsec:numa_migrations}

When an OS-level NUMA allocation policy is not specified,
the OS can dynamically migrate data among NUMA nodes to increase local NUMA
accesses. NUMA page migrations are helpful for multiple applications sharing a
single system as they try to move pages closer to the cores assigned to each
application. However, for a single application,
this policy may not always be useful, especially when application has specified
its own allocation policy.

\begin{figure*}[h]
    \small
    \centering
    \begin{minipage}{0.63\textwidth}
    \centering
    \includegraphics[width=\textwidth]{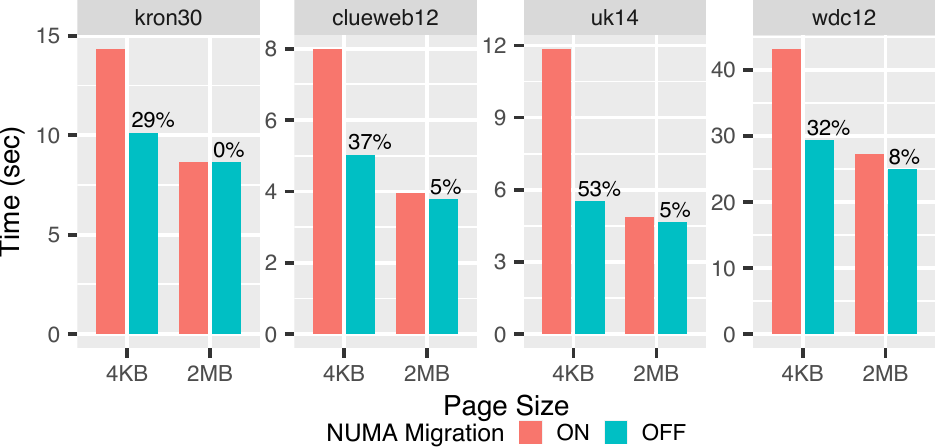}
    (a) \xpointshort
    \end{minipage}
    \begin{minipage}{0.33\textwidth}
    \centering
    \includegraphics[width=\textwidth]{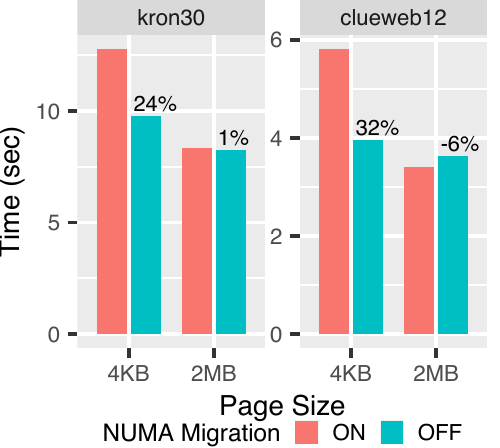}
    (b) DDR4 DRAM
    \end{minipage}
    \caption{Execution time of bfs in Galois using small (4KB) and huge (2MB) page sizes with and without NUMA migration.}
    \label{fig:res:numa_migration_benchmark}
    \vspace{-5pt}
\end{figure*}

\begin{figure*}
    \small
    \centering
    \begin{minipage}{\textwidth}
    \begin{minipage}{0.2\textwidth}
    \centering
    \includegraphics[width=\textwidth]{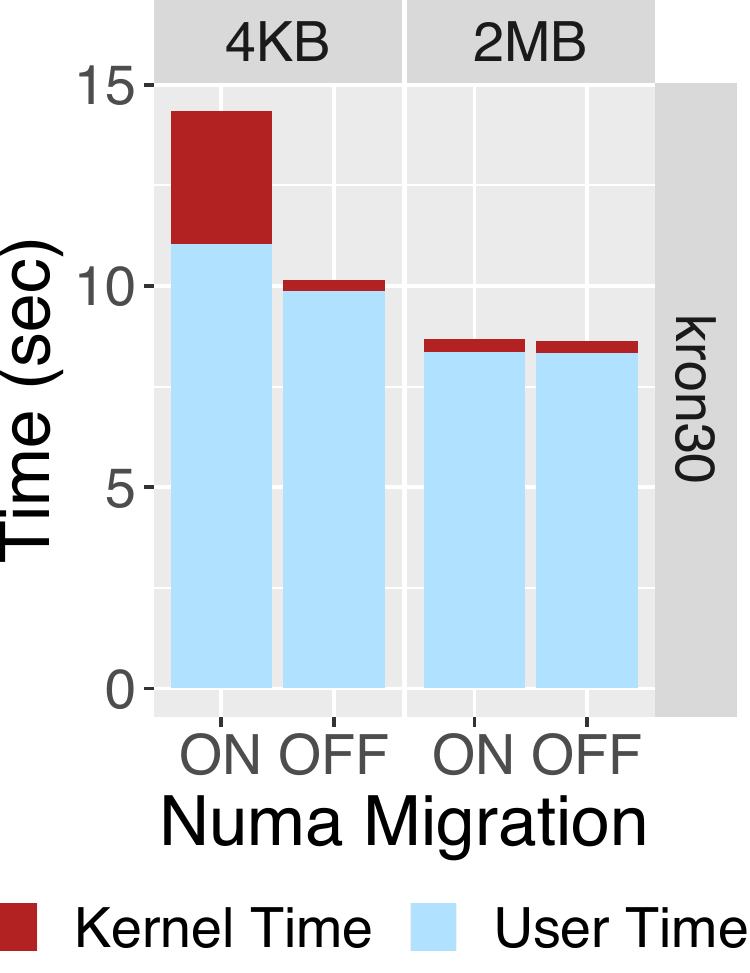}
    (a) \xpointshortend
    \end{minipage}
    \hspace{10pt}
    \begin{minipage}{0.2\textwidth}
    \centering
    \includegraphics[width=\textwidth]{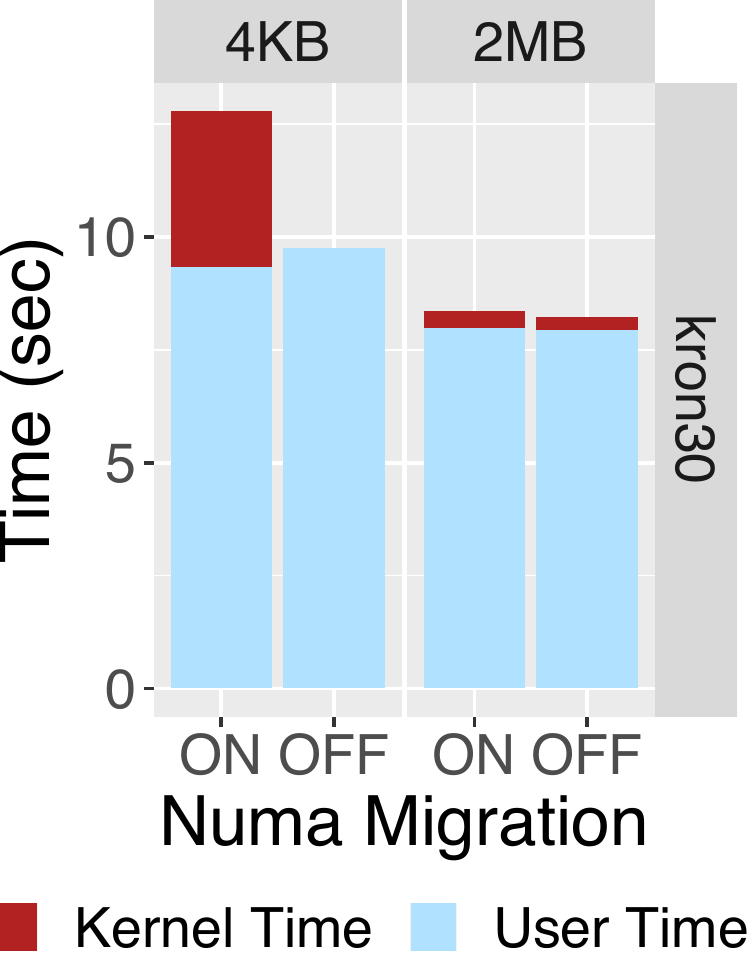}
    (b) DDR4 DRAM
    \end{minipage}
    \hspace{30pt}
    \vline
    \hspace{30pt}
    \begin{minipage}{0.2\textwidth}
    \centering
    \includegraphics[width=\textwidth]{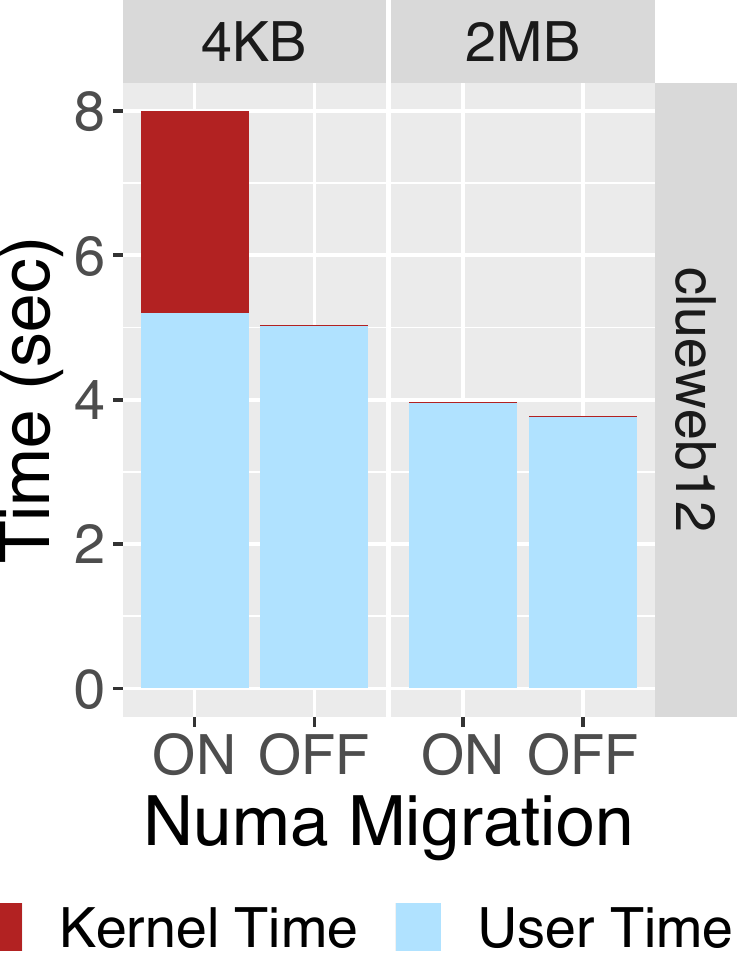}
    (a) \xpointshortend
    \end{minipage}
    \hspace{10pt}
    \begin{minipage}{0.2\textwidth}
    \centering
    \includegraphics[width=\textwidth]{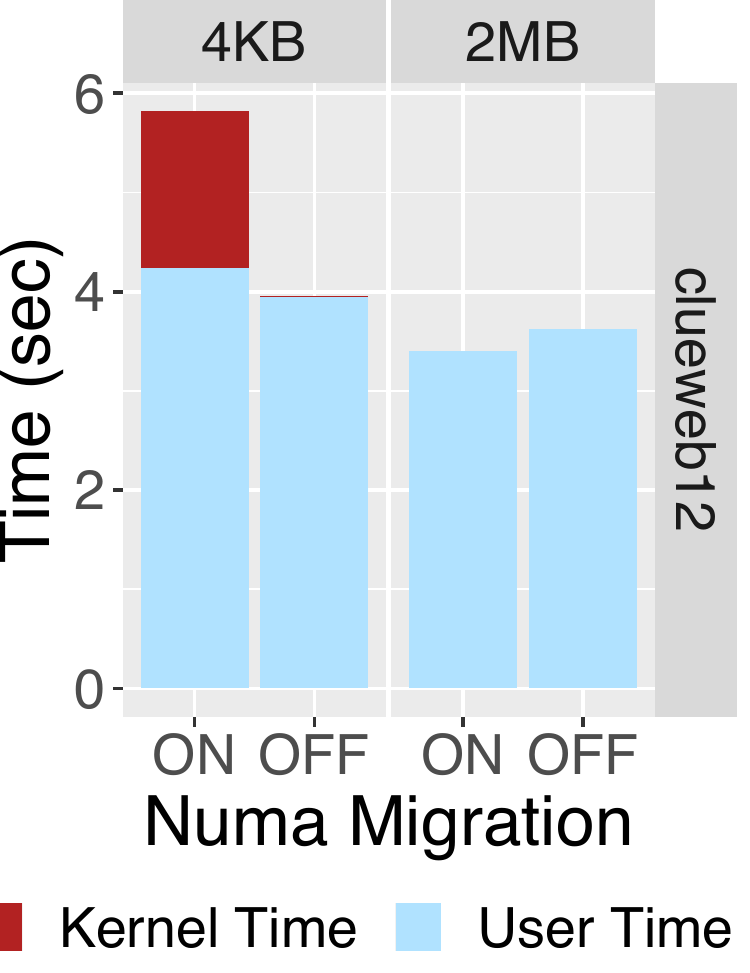}
    (b) DDR4 DRAM
    \end{minipage}
    \end{minipage}
    \caption{Breakdown of execution time of bfs in Galois 
    using different page sizes 
    for kron30 (left) and clueweb12 (right).
    }
    \label{fig:res:numa_migration_benchmark_breakdown}
\end{figure*}
NUMA migration has overheads: (a) it requires book-keeping to
track accesses to the pages to select pages for migration, and
(b) migration changes the virtual-to-physical address mapping, which makes
the Page Table Entries (PTEs) cached in CPU's Translation
Lookaside Buffers (TLBs)
stale, causing TLB shootdown on each core \revise{to invalidate stale entries}.
TLB shootdown involves slow operations such as issuing
inter-processor-interrupts (IPIs), and it also increases TLB misses.

Graph analytics applications tend to
have irregular access patterns: accesses are arbitrary, so there may be many
shared accesses across NUMA nodes.
To examine the effects of page migration on graph analytics applications, we
run breadth-first search (bfs)
(similar trends observed for other
benchmarks) with Galois~\cite{galois} using NUMA interleaved allocation
on both \xpointshort and DDR4 DRAM with NUMA migration on and off.
We also examine the effects of page migration for different page sizes
\revise{(which affects the number of pages migrated)}: (a) 4KB small pages
and (b) 2MB huge pages.  The results suggest that \emph{NUMA migration should
be turned off for graph analytics applications on \xpointshortend}.


Figure~\ref{fig:res:numa_migration_benchmark} shows the effect of NUMA
migration where the number on each bar presents the \% change in the
execution time when NUMA migration is turned off. A positive number means
turning migration off improves performance. Performance improves in most
cases if migration is turned off.
Figure~\ref{fig:res:numa_migration_benchmark_breakdown} shows that the
time spent in user code is not affected by the migrations, which shows
that they add kernel time overhead without giving significant
benefits.  Another way to measure the efficacy of the migrations is to
measure the \% of local near-memory (DRAM) accesses in
\xpointshortend: if migration is beneficial, then it should increase.
However, this does not change by more than $1\%$.
Figure~\ref{fig:res:numa_migration_benchmark_breakdown} shows that 
migrations hurt performance more on \xpointshort compared to DRAM as 
kernel time spent is higher. This is due to (a) higher bookkeeping cost as
accesses to kernel data structures are more expensive on Optane and (b)
higher cost of TLB shootdown
as it increases the access latency to the near-memory (DRAM) being used as a
direct-mapped cache since TLB translation is on the critical
path.\footnote{Since near-memory is physically indexed and
physically tagged direct-mapped cache, virtual addresses are 
translated to physical addresses before cache (near-memory) can be
accessed.\label{fn:tlb}}. Larger graphs exacerbate this effect as they use
more pages.

4KB small page size shows more performance improvement than 2MB huge pages
when turning page migrations off.
%
We observe that the number of migrations 
is in the millions for small pages and in the hundreds for
huge pages. The finer granularity of small pages makes them more prone to
migrations, leading to more TLB shootdowns and data TLB misses. Therefore,
for small pages, the number of data TLB misses 
reduces by $\sim2\times$ by
turning off the migrations for all the graphs.
The number of small pages being 512$\times$
the number of huge pages also increases the bookkeeping overhead in the OS.
This is reflected in time spent in the OS kernel
seen in Figure~\ref{fig:res:numa_migration_benchmark_breakdown}:
the time spent in the kernel is more for the smaller page size than for the
larger page size if migration is turned on. 



\subsection{Page Size Selection}
\label{alloc-page-size}
\label{subsubsec:hupe_vs_small}

When memory sizes and workload sizes grow, the time spent handling TLB misses
can become a performance bottleneck since large working sets need many
virtual-to-physical address translations that may not be cached in the TLB.
This bottleneck can be tackled (a) by increasing the TLB size 
or (b) by increasing the page size. The TLB size is determined by
the micro-architecture and cannot easily be changed by a user. On the other
hand, processors allow users to customize page sizes as different
page sizes may work best for different workloads.
For example, x86 supports traditional 4KB small pages as well as 2MB and 1GB
huge pages.

We studied the impact of page size on graph analytics using a 4KB small page
size and a 2MB huge page size. We did not include the 1GB page size as
it requires special setup; moreover, we do not expect to gain from 1GB
pages as the hardware supports fewer TLB entries for 1GB page size. We run
bfs (similar behavior observed for other benchmarks) using
Galois~\cite{galois} with no NUMA migration and NUMA interleaved 
allocation for various large graphs on (a) \xpointshort and (b) DDR4
DRAM. The results suggest that \emph{a page size of 2MB is
good for graph analytics on \xpointshortend}.

Figure~\ref{fig:res:numa_migration_benchmark} shows bfs runtimes with
various page sizes, and we observe that using huge pages is always beneficial
on large graphs as huge pages reduce the number of pages
by 512$\times$, reducing the number of TLB misses (3.2$\times$
for clueweb12, 11.2$\times$ for uk14 and 1.9$\times$ for wdc12) 
and CPU cycles spent on page walking on
TLB misses (7.3$\times$ for clueweb12, 12.5$\times$ for uk14 and 8.8$\times$
for wdc12).
We also observe that the benefits of huge pages are higher on \xpointshort than
on DRAM because TLB misses increase the near-memory access latency. Huge pages
increase the TLB reach (TLB size $\times$ page size), thereby reducing the TLB
misses.

\subsection{Summary}

For high-performance graph analytics on \xpointshortend, we recommend
(i) NUMA interleaved or blocked allocation rather than NUMA
local, particularly for large allocations ($>$ 192GB), (ii) turning off NUMA page
migration, and (iii) using 2MB huge pages.

%% file: operator.tex
\begin{figure*}[ht]
    \footnotesize
    \centering
    \begin{minipage}{\textwidth}

    \begin{minipage}{0.32\textwidth}
        \centering
    \includegraphics[width=\textwidth]{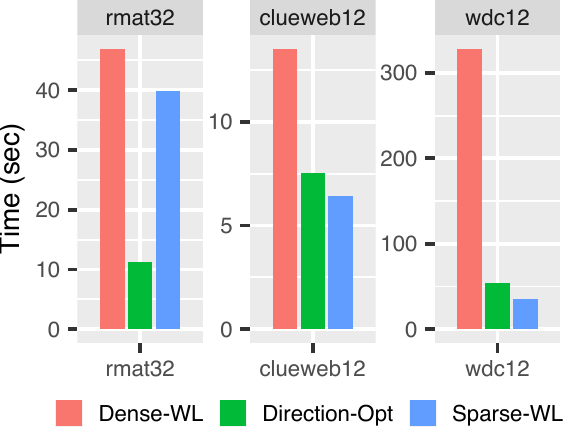}
    (a) bfs
    \end{minipage}
    \begin{minipage}{0.32\textwidth}
    \centering
    \includegraphics[width=\textwidth]{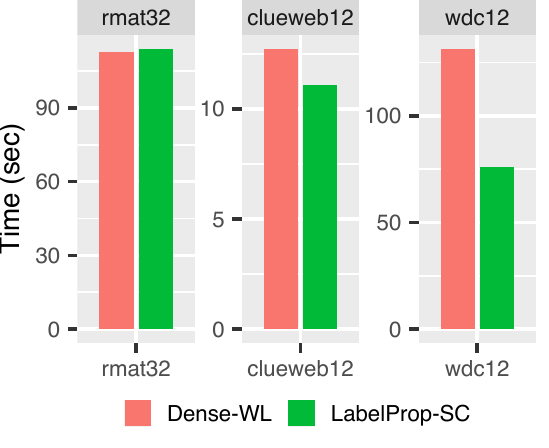}
    (b) cc
    \end{minipage}
   \begin{minipage}{0.32\textwidth}
    \centering
    \includegraphics[width=\textwidth]{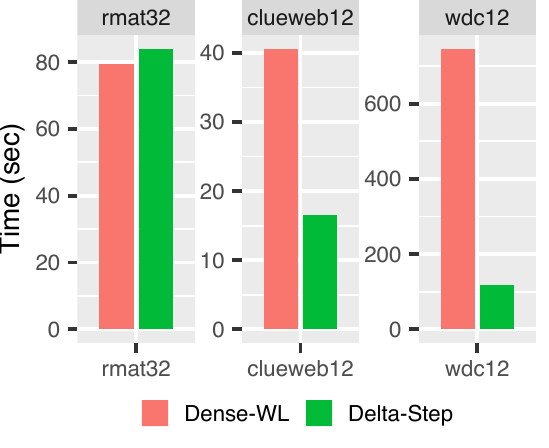}
    (c) sssp
    \end{minipage}

    \end{minipage}
    \caption{Execution time of different data-driven algorithms in Galois on
    \xpointshort using 96 threads.}
    \label{fig:res:algo_variant_plot}
\end{figure*}

\begin{figure*}[ht]
    \footnotesize
    \centering
    \begin{minipage}{\textwidth}

    \begin{minipage}{0.32\textwidth}
        \centering
    \includegraphics[width=\textwidth]{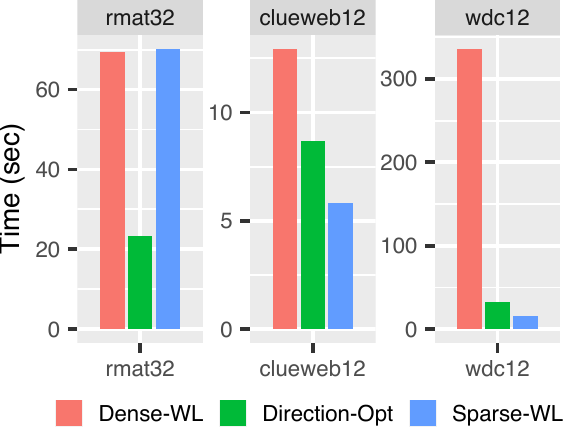}
    (a) bfs
    \end{minipage}
    \begin{minipage}{0.32\textwidth}
    \centering
    \includegraphics[width=\textwidth]{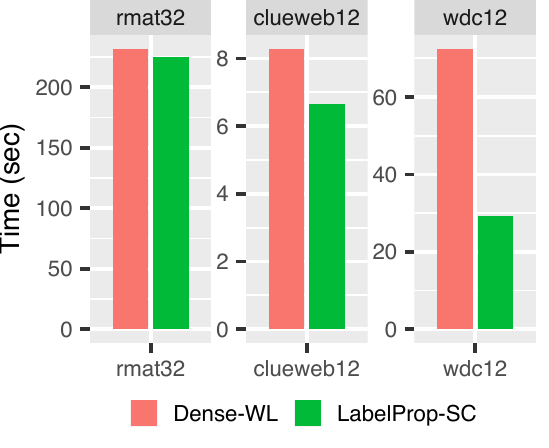}
    (b) cc
    \end{minipage}
    \begin{minipage}{0.32\textwidth}
    \centering
    \includegraphics[width=\textwidth]{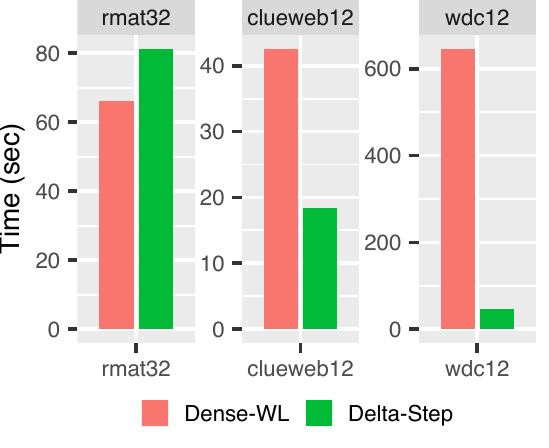}
    (c) sssp
    \end{minipage}
    \end{minipage}
    \caption{Execution time of different data-driven algorithms in Galois on
    Entropy (1.5TB DDR4 DRAM) using 56 threads.}
    \label{fig:res:algo_variant_plot_dram}
\end{figure*}


Generally, there are many algorithms that can solve a given graph analytics
problem; for example, \revise{the single-source shortest-path (sssp) problem
can be solved using Dijkstra's algorithm~\cite{cormen01}, the Bellman-Ford algorithm~\cite{cormen01}, chaotic
relaxation~\cite{chaoticrelax}, or delta-stepping~\cite{meyer98}}. These algorithms may have different
asymptotic complexities and different amounts of parallelism. For a graph
$G=(V,E)$, the asymptotic complexity of Dijkstra's algorithm is
$O(|E|{*}log(|V|))$ while Bellman-Ford is $O(|E|{*}|V|)$, but for most graphs,
Dijkstra's algorithm has little parallelism compared to Bellman-Ford.
Complicating matters further is the fact that a given algorithm can usually be
implemented in different ways that can affect parallel performance
dramatically; implementations with fine-grain locking, for example, usually
perform better than those with coarse-grain locking.

This section presents a classification of graph analytics algorithms that is
useful for understanding parallel performance~\cite{pingali11}. We also present
experimental results that provide insights into which classes of algorithms
perform well on large input graphs.

\subsection{Classification of Graph Analytics Algorithms}
\label{subsec:class-graph}

\noindent{\bf Operators}
In graph analytics algorithms, each vertex has one or more labels that are
initialized at the start of the computation and updated repeatedly during
computation until a quiescence condition is reached. Label updates are
performed by applying an {\em operator} to \emph{active vertices} in the
graph. In some systems, such as Galois~\cite{galois}, an
operator may read and update an arbitrary portion of the graph surrounding
the active vertex; this portion is called its {\em neighborhood}. \revise{Most
shared-memory systems such as Ligra~\cite{ligra,gbbs} and
GraphIt~\cite{graphit} only support {\em vertex programs} in which operator 
neighborhoods are only the immediate 
neighbors of the active vertex.  The more general non-vertex programs, 
conversely, have no restriction on an operator's neighborhood.}
A {\em push-style} operator updates the labels of the neighbors of the active
vertex, while a \emph{pull-style} operator updates the label of only the active
vertex.  {\em Direction-optimizing} implementations~\cite{dobfs} can switch
between push and pull style operators dynamically but require a reverse
edge for every forward edge in the graph, doubling the memory footprint of
the graph.

\noindent{\bf Schedule}
\revise{To find active vertices in the graph, algorithms take one of two approaches. A
{\em topology-driven} algorithm executes in rounds. In each round, it
applies the operator to all vertices; Bellman-Ford sssp is an example. These
algorithms are simple to implement, but they may not be work-efficient if
there are few active vertices in a lot of rounds. To address this, {\em
data-driven} algorithms track active vertices explicitly and only apply the operator
to these vertices. At the start of the algorithm, some vertices are active;
applying the operator to an active vertex may activate other vertices, and
operator application continues until there are no active vertices in the graph.
Dijkstra and delta-stepping sssp algorithms are examples.
Active vertices can be tracked using a bit-vector of size $V$ if there are $V$
vertices in the graph: we call this a {\em dense
worklist}~\cite{ligra,gbbs,graphit}. Other implementations keep an explicit
worklist of active vertices~\cite{galois}: we call this a {\em sparse
worklist}.}

Some implementations of data-driven algorithms execute in {\it
bulk\hyp{}synchronous} rounds: they keep a \emph{current} and a \emph{next}
worklist, and in each round, they process only vertices in the current
worklist and add activated vertices to the next worklist. The worklists can be
dense or sparse. In contrast, {\em asynchronous} data-driven implementations
have no notion of rounds; they maintain a single sparse worklist, pushing and
popping active vertices from this worklist until it is empty.

\subsection{Algorithms for Very Large Graphs}
\label{subsec:algo_large_graphs}

At present, very large graphs are analyzed using clusters or out-of-core
systems, but these systems are restricted to vertex programs
and round-based execution. This is not considered a serious limitation
for power-law graphs since they have a small diameter and information does not
have to propagate many hops in these graphs.  In fact, no graph analytics
framework other than Galois provides sparse worklists, so they do not support
asynchronous data-driven algorithms, and most of them are restricted to vertex
programs.

Using \xpointshortend, we use a single machine to
perform analytics on very large graphs, and our results suggest that
conventional wisdom in this area needs to be revised. {\em The key issue is
highlighted by Table~\ref{tbl:inputs}: clueweb12, uk14, and wdc12, which are
real-world web-crawls, have a high diameter (shown in Table~\ref{tbl:inputs})
compared to kron30 and rmat32, the synthetic power-law graphs.}
We show that for standard graph analytics problems, the best-performing
algorithms for these graphs may be (a) non-vertex programs and (b)
asynchronous data-driven algorithms, which require sparse worklists.  These
algorithms have better work-efficiency and make fewer memory accesses, which is
good for performance especially on \xpointshort where memory accesses are
more expensive.

\begin{figure*}
    \centering
    \begin{minipage}{0.49\textwidth}
      \centering
        \includegraphics[width=\textwidth]{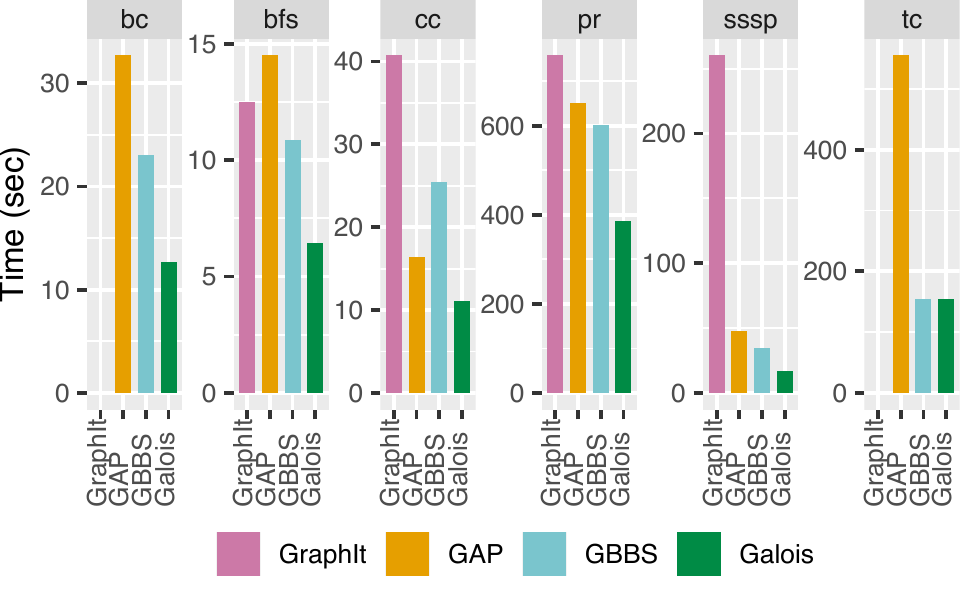}
        (a) clueweb12
        \vspace{8pt}
    \end{minipage}
    \begin{minipage}{0.49\textwidth}
      \centering
        \includegraphics[width=\textwidth]{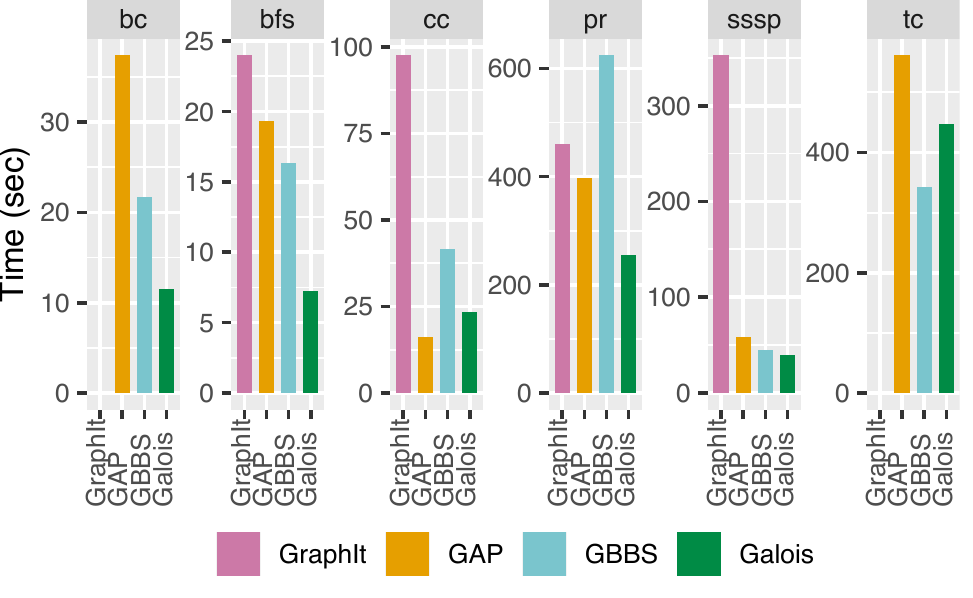}
        (b) uk14
        \vspace{8pt}
    \end{minipage}
    \begin{minipage}{0.49\textwidth}
      \centering
        \includegraphics[width=\textwidth]{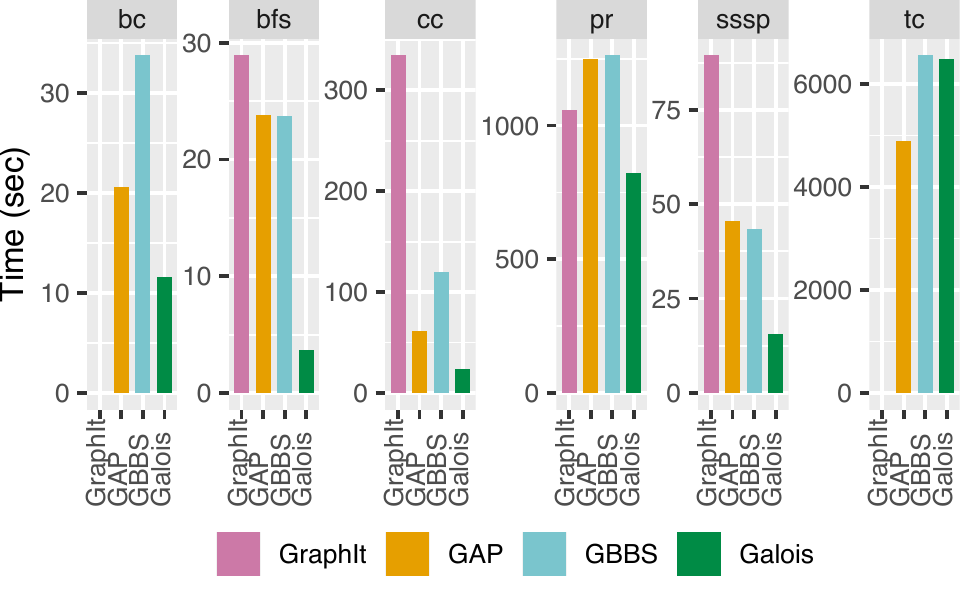}
        (c) \revise{iso\_m100}
    \end{minipage}
    \begin{minipage}{0.49\textwidth}
      \centering
        \includegraphics[width=\textwidth]{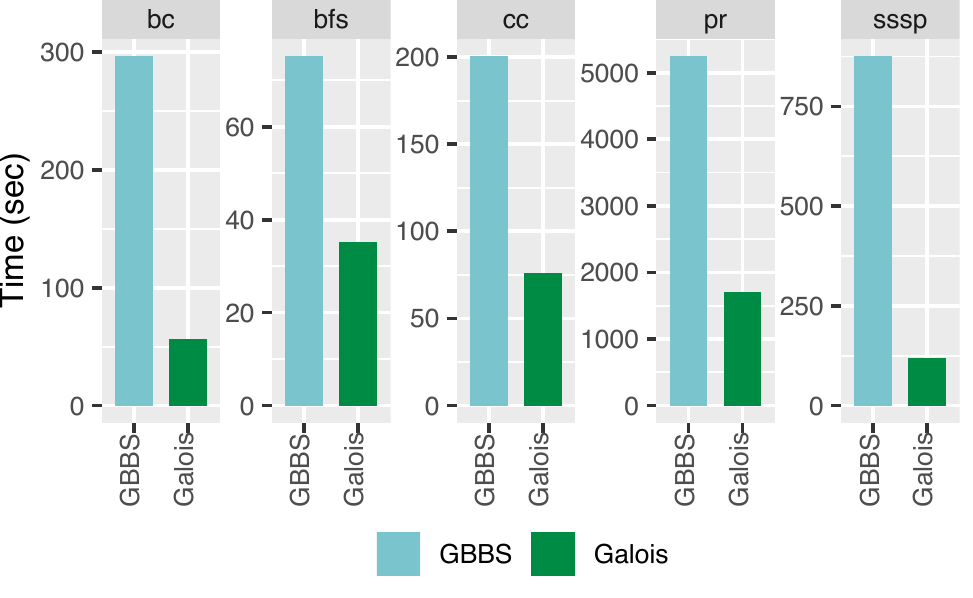}
        (d) wdc12
    \end{minipage}
    \caption{Execution time of benchmarks in GraphIt, GAP, GBBS, and Galois
    on \xpointshort using 96 threads.}
    \label{fig:res:other_vs_galois}
    \vspace{-5pt}
    \end{figure*}

    

Figure~\ref{fig:res:algo_variant_plot} shows the execution time of different
data-driven algorithms for bfs, cc, and sssp on \xpointshort using the rmat32,
clueweb12, and wdc12 graphs on the Galois system.  For bfs, all algorithms are
bulk\hyp{}synchronous.  A vertex program with direction optimization (that uses
dense worklists) performs well for rmat32 since it has a low-diameter, but for
the real-world web-crawls which have higher diameter, it is outperformed
by an implementation with a push-style operator and sparse worklists since this
algorithm has a lower memory footprint and makes fewer memory accesses.
For cc, bulk\hyp{}synchronous \revise{label propagation~\cite{lpcc}} combined
with short-cutting
(LabelProp-SC)~\cite{shortcutcc},
which uses a non-vertex operator, is used. It is a variant of the
\revise{Pointer-Jumping algorithm~\cite{pjcc}} where after a round of
label propagation it jumps one level unlike Pointer-Jumping where it goes
to the common ancestor. LabelProp-SC exhibits better locality 
compared Pointer-Jumping and significantly outperforms the
bulk\hyp{}synchronous algorithm that uses a simple label propagation vertex
operator for the real-world web-crawls. For sssp, the asynchronous
delta-stepping algorithm, which maintains a sparse worklist, significantly
outperforms the bulk\hyp{}synchronous data-driven algorithm with dense
worklists.
These findings do not apply only to \xpointshortend:
Figure~\ref{fig:res:algo_variant_plot_dram} shows the same experiments for
bfs, sssp, and cc conducted on Entropy (DDR4 DRAM machine).
The trends are similar to those on \xpointshort machine.

\noindent{\bf Summary}
Large real-world web-crawls, which are the largest graphs available today,
have a high diameter unlike synthetically generated rmat and kron
graphs. Therefore, conclusions drawn from experiments with rmat and kron
graphs can be misleading. On current distributed-memory and out-of-core
platforms, one is forced to use vertex programs, but on machines with
\xpointshortend, \emph{it is advantageous to use algorithms with non-vertex
operators and sparse worklists of active vertices that allow for
asynchronous execution}. Frameworks with only vertex
operators or no sparse worklists are at a disadvantage on this
platform when processing large real-world web-crawls, as we show next.

%% file: results.tex
In this section, we evaluate several graph frameworks on \xpointshort in the
context of the performance guidelines presented in
Sections~\ref{subsec:3dxpoint_setup} and~\ref{subsec:algo_design}.  In
Section~\ref{subsec:framework_comparison}, four shared-memory graph analytics
systems - Galois~\cite{galois}, GAP~\cite{gap}, GraphIt~\cite{graphit}, and
GBBS~\cite{gbbs} - are evaluated on the \xpointshort machine using several
graph analytics applications.
Section~\ref{subsec:xpoint_vs_dram} describes experiments with medium-sized
graphs stored either in \xpointshort or DRAM. These experiments provide
end-to-end estimates of the overhead of executing applications with data in
\xpointshort rather than in DRAM.  Section~\ref{subsec:xpoint_vs_stampeded}
describes experiments with large graphs that fit only in \xpointshortend, and
performance is compared with distributed-memory performance on a production
cluster with up to 128 machines. Section~\ref{subsec:outofcore} presents
our experiments with GridGraph~\cite{gridgraph}, an out-of-core graph
analytics framework, using \xpointshortend's app-direct mode to
treat it as external memory.


\subsection{Galois, GAP and GraphIt on \xpointshortend}
\label{subsec:framework_comparison}
\input{results_3rdparty}

\subsection{Medium-size graphs: Using \xpointshort vs. DDR4 DRAM}
\label{subsec:xpoint_vs_dram}
\input{DRAMvsXPoint.tex}

\subsection{Very large graphs: Using \xpointshort vs. a Cluster}
\label{subsec:xpoint_vs_stampeded}
\input{DistributedvsXPoint.tex}

\subsection{Out-of-core GridGraph in App-direct Mode vs. Galois in
Memory Mode}
\label{subsec:outofcore}
\input{outofcore.tex}

\subsection{Summary and Discussion}
\label{subsec:summary}
\input{results_summary}

%% file: results_3rdparty.tex
\begin{figure*}
  \centering
  \begin{minipage}{0.99\textwidth}
  \centering
  \includegraphics[width=0.99\textwidth]{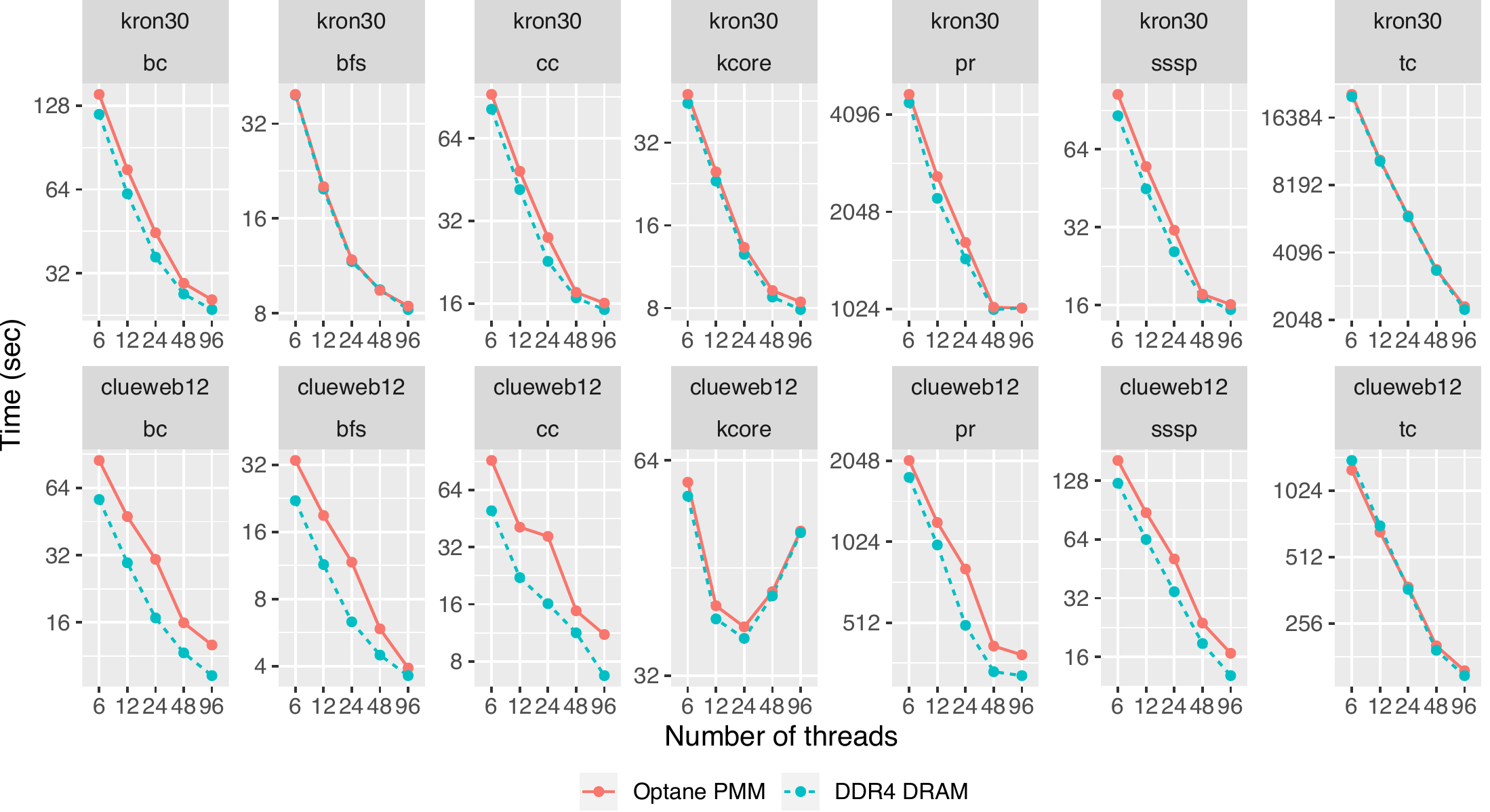}
  \end{minipage}
  \caption{Strong scaling in execution time of benchmarks in Galois using DDR4 DRAM and \xpointshortend.}
  \label{fig:res:dram_vs_xpoint}
  \end{figure*}
  
\noindent{\bf Setup.}
\revise{To choose a shared-memory graph analytics system for our experiments, we
evaluate (1) Galois~\cite{galois}, which is a library and runtime for graph
processing, (2) GAP~\cite{gap}, which is a benchmark suite of expert-written graph
applications, (3) GraphIt~\cite{graphit}, which is a domain-specific
language (DSL) and optimizing compiler for graph computations, and
(4) GBBS~\cite{gbbs}, which is a benchmark suite of graph algorithms 
written in the Ligra~\cite{ligra} framework.}

\revise{The choice of these frameworks was made as they exemplify different
approaches to shared-memory graph analytics. GraphIt is a DSL that only
supports vertex programs, and it has a compiler that uses auto-tuning to
generate optimized code; the optimizations are controlled by the
programmer.  Galois is a C++-based general-purpose programming system based on
a runtime that permits optimizations to be specified in the program at
compile-time or at runtime, giving the application programmer a large design
space of implementations that can be explored.  GBBS programs are expressed in
a graph processing library and runtime, Ligra. Therefore, Galois and GBBS
require more programming effort than GraphIt.  GBBS includes theoretically
efficient algorithms written by experts, while 
Galois includes algorithms written using 
expert-provided concurrent data
structures and operator schedulers.
GAP is a benchmark suite of graph analytics applications written by expert
programmers; it does not provide a runtime or data structures like Galois.} 

The kcore application is not implemented in GAP and GraphIt,
so we omit it in the comparisons reported in this section.
We omit the largest graph wdc12
for GAP and GraphIt because neither can handle graphs
that have more than $2^{31}{-}1$ nodes (they use a
\texttt{signed 32-bit int} for storing node IDs).
GAP, GraphIt, and GBBS do not use NUMA allocation policies within their applications,
so we use the OS utility \textsf{numactl} to use NUMA interleaved.
For Galois, we use the best-performing algorithm with a runtime option, and
we did not try different worklists or chunk-sizes.
Galois allows programmers to choose NUMA interleaved or blocked
allocation for each application by modifying a template argument in
the program; we choose interleaved for bfs, cc, and sssp and blocked for bc,
and pr. 
For GraphIt, we used
the optimizations recommended by the authors~\cite{graphit} in the GraphIt
artifact.

\noindent{\bf Results.}
Figures~\ref{fig:res:other_vs_galois} shows the
execution times on \xpointshort (GraphIt does not have bc).
Galois is generally much faster than GraphIt, GAP, and GBBS: on average,
Galois is $3.8\times$, $1.9\times$, and $1.6\times$ faster than GraphIt, GAP,
and GBBS, respectively.
There are many reasons for these performance differences.

Algorithms and implementation choices affect runtime (discussed
in Section~\ref{subsec:algo_large_graphs}). 
For all algorithms, GAP, GBBS and GraphIt use a dense worklist to store the
\textit{frontier} while Galois uses a sparse worklist except for pr
(large diameter graphs tend to have sparse frontiers).
All systems use the same algorithm for pr.  For bfs, all systems except Galois 
use direction-optimization which accesses both in-edges and out-edges (increasing
memory accesses). 
For sssp, GAP, GBBS, and Galois use \revise{delta-stepping~\cite{meyer98}}; 
GraphIt does not support such algorithms. 
For cc, GAP and GBBS use a union-find based \revise{\textit{pointer-jumping}~\cite{pjcc}
algorithm, Galois uses label-propagation with shortcutting~\cite{shortcutcc}, and GraphIt uses
label propagation~\cite{lpcc}} because it supports only vertex programs.
Furthermore, Galois uses asynchronous execution for sssp and cc unlike the
others.

Another key difference is how the three systems perform memory
allocations. Galois is the only framework that explicitly uses huge pages of
size 2MB whereas the others use small pages of size 4KB and rely on
the OS to use {\em Transparent Huge Pages (THP)}. As discussed in
Section~\ref{subsec:3dxpoint_setup}, huge pages can significantly reduce the
cost of memory accesses over small pages even when THP is enabled.  Galois is
also the only one to provide NUMA blocked allocation, and we chose that policy
because it performed observably better than the interleaved policy for some benchmarks such
as bc and pr (the performance difference was within 18\%). In general, we observe that 
NUMA blocked performs better for topology-driven algorithms, while NUMA interleaved 
performs better for data-driven algorithms.
%
In addition, GAP, GBBS, and GraphIt allocate memory for both in- and
out-edges of the graph while Galois only allocates memory only for the
direction(s) needed by the algorithm. Allocating both increases the memory
footprint and leads to conflict misses in near-memory when both in-edges and
out-edges are accessed.

\revise{To conclude, our experiments show that in order to acheive performance
for large graphs, a framework should support asynchronous, non-vertex programs
as well as allow users explicit control over memory allocation.  As Galois
supports these features out-of-the-box, we use it for the rest of our experiments.}

%% file: DRAMvsXPoint.tex
\noindent{\bf Setup.}
\revise{In this subsection, we determine the overhead of using \xpointshort
over DRAM by examining the runtimes for graphs that are small enough to fit in
DRAM (384 GB).}  We use with kron30 and clueweb12 (Table~\ref{tbl:inputs})
which both fit in DRAM.  We use algorithms in Galois that perform best on 96
threads. 

\noindent{\bf Results.}
Figure~\ref{fig:res:dram_vs_xpoint} shows strong scaling on DRAM
and on \xpointshort with DRAM as cache.
kron30 requires $\sim136$GB, which is a third of the near-memory available, so
\xpointshort is almost identical to DRAM as it can cache the graph
in near-memory effectively. On the other hand, clueweb12 requires $\sim365$GB, which
is close to the near-memory available, so there are significantly more
conflict-misses ($\approx26\%$) in near-memory.  On 96
threads, \xpointshort can take up to 65\% more execution time than DRAM, but on
average, it takes only $7.3\%$ more time than DRAM.

Another trend is that if the number of threads is less than 24, \xpointshort
can be slower than DRAM because of the way Galois allocates memory.
Interleaved and blocked allocation policies in Galois interleave and block
among \emph{threads}, not sockets.  If threads used is less than
24, all threads run on one socket, and all memory is allocated there, leading
to under-utilization of the DRAM in the entire system: this results in more
conflict-misses in near-memory. 

\revise{The strong scaling for all the applications is similar whether 
DRAM or Optane is used as main memory.}
\revise{We also measured the performance of all applications 
for larger graphs on 8 
and 96 threads of Optane. 
The geo-mean speedup of all applications 
on 96 threads over 8 threads 
is $4.3\times$, $4.2\times$, $4.7\times$, and $3.5\times$ 
for kron30, clueweb12, uk14, and wdc12 respectively. 
Thus, the strong scaling speedup does not vary by much 
as the size of the graph grows.}


%% file: DistributedvsXPoint.tex
\input{tbl_stampede_vs_optane}

\noindent{\textbf{Setup.}}
For very large graphs that do not fit in DRAM, the conventional choices are to
use a distributed or an out-of-core system. \revise{In this subsection, we
compare execution of Galois on \xpointshort to execution of the
state-of-the-art distributed graph analytics system D-Galois~\cite{gluon} on
the Stampede2~\cite{stampede} cluster to determine the competitiveness of
\xpointshort compared to a distributed cluster. 
We chose D-Galois because it is 
faster than existing distributed graph systems 
(like Gemini~\cite{gemini}, PowerGraph~\cite{powergraph}, 
and GraphX~\cite{graphx})
and on each
machine, it uses the same computation runtime as Galois, 
which makes the comparison fair.} To
partition graphs between machines, we follow the recommendations of a previous
study~\cite{partitioningstudy} and use Outgoing Edge Cut (OEC) for 5 and 20
hosts and Cartesian Vertex Cut (CVC)~\cite{cvc2d} for 256 hosts.  
D-Galois supports only bulk-synchronous vertex programs with dense worklists as
they simplify communication.  Therefore, it cannot support some of the more
efficient non-vertex programs in Galois.  We exclude graph loading,
partitioning, and construction time in the reported numbers.
\revise{For logistical reasons, it is difficult to ensure that both platforms
use the exact same resources (threads and memory). 
For a fair comparison, 
we limit the resources used on both platforms.}




\noindent{\textbf{Results.}} Table~\ref{tbl:stampvsoptane} compares the
performance of \xpointshort (on a single machine) running non-vertex,
asynchronous Galois programs (referred to as \textbf{OB}) with the distributed cluster
(Stampede2~\cite{stampede}) running D-Galois vertex programs using the
minimum number of hosts required to hold the graph in memory (5 hosts for
clueweb12, and uk14, and 20 hosts for wdc12; 48 threads per host and referred
to as \textbf{DM}). \xpointshort outperforms D-Galois in most
cases (best times highlighted), except for pr on clueweb12, uk14,
and wdc12 and cc on uk14, and wdc12. \xpointshort gives a geomean speedup of
$1.7\times$ over D-Galois even though D-Galois has more cores (240 cores
for clueweb12 and uk14; 960 cores for wdc12) and memory bandwidth. In pr,
almost all nodes are updated in every round (similar to the topology driven
algorithms); therefore, on the distributed cluster, it benefits from
the better spatial locality in D-Galois resulting from the partitioning of the
graph into smaller local graphs and more memory bandwidth.

\revise{Table~\ref{tbl:stampvsoptane} also shows the performance 
of many applications on Optane 
for input graphs of different sizes (smallest to largest). 
Most applications take more time for larger graphs. 
Although wdc12 is $\sim3\times$ larger (in terms of edges) 
than clueweb12, 
kcore and pr take less time on wdc12 than clueweb12, 
but the other applications take on average $\sim7\times$ 
more time on wdc12 than clueweb12. 
This provides an estimate for the weak scaling of these 
applications.}

\begin{figure*}
    \begin{minipage}{0.99\textwidth}
    \includegraphics[width=0.99\textwidth]{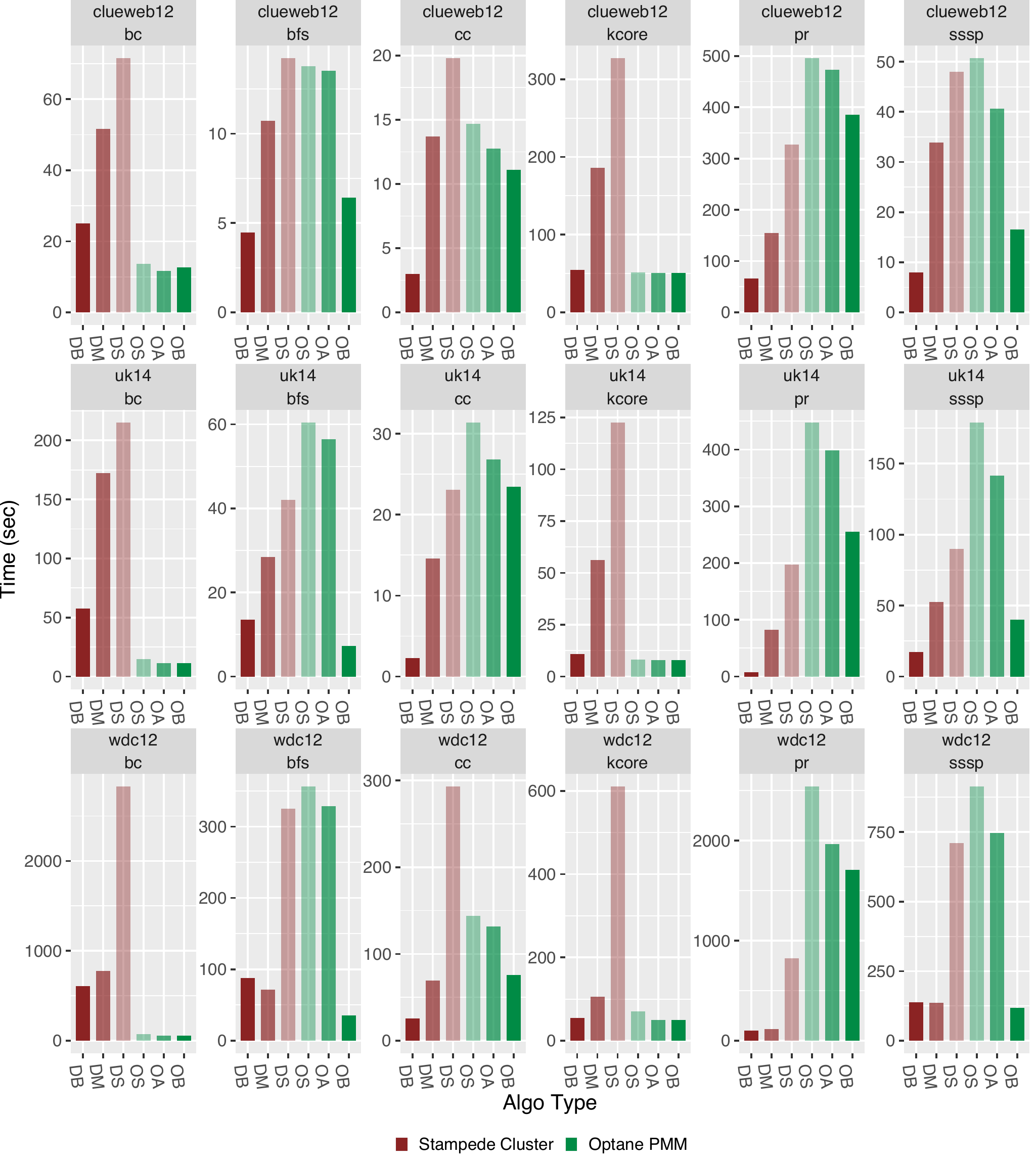}
    \caption{Execution time of benchmarks in Galois on \xpointshort machine and
    D-Galois on Stampede cluster
    with different configurations
    :-
    DB: Distributed Best (all threads on 256 hosts),
    DM: Distributed Min (all threads on min \#hosts that hold graph),
    DS: Distributed Same (total 80 threads on min \#hosts that hold graph),
    OS: Optane Same (same algorithm and threads as DS),
    OA: Optane All (same algorithm as DS, DM, and DB on 96 threads),
    OB: Optane Best (best algorithm on 96 threads).}
    \label{fig:res:stampede_vs_xpoint_same_cores}
    \end{minipage}
    \end{figure*}

\noindent{\textbf{Further Analysis.}}
The bars labeled \textbf{O\_} in Figure~\ref{fig:res:stampede_vs_xpoint_same_cores}
show times on the \xpointshort system with the following configurations:-
%
\textbf{OB}: Performance using the best algorithm in Galois for that problem and all 96 threads 
(same as shown in Table~\ref{tbl:stampvsoptane});
%
\textbf{OA}: Performance using the best vertex programs in Galois for that problem and all 96 threads;
%
\textbf{OS}: Same as \textbf{OA} but using only 80 threads.
%
The bars labeled \textbf{D\_} show times on the Stampede2 system with the following configurations:-
%
\textbf{DB}: Performance using D-Galois vertex programs on 256 machines (12,288 threads);
%
\textbf{DM}: Performance using D-Galois vertex programs using the minimum number
of hosts required to hold graph in memory (same as shown in Table~\ref{tbl:stampvsoptane}).
%
\textbf{DS}: Same as \textbf{DM} but using a total of 80 threads across all machines.

\noindent{\textbf{Results.}}
Figure~\ref{fig:res:stampede_vs_xpoint_same_cores} shows the experimental
results.
%
For bars \textbf{DS} and \textbf{OS}, the algorithm and resources are roughly
the same, so in most cases, \textbf{OS} is similar or better than
\textbf{DS}. The notable exception to this is pr (reason is explained above).
On average, \textbf{OS} is 1.9$\times$ faster than \textbf{DS} for all
inputs and benchmarks.
Bars \textbf{OB} and \textbf{OA} show the advantages of using non-vertex,
asynchronous programs on the \xpointshort system. Bars \textbf{DB} and
\textbf{OB} show that with the more complex algorithms that can be
implemented on the \xpointshort system, performance on this system matches
the performance of vertex programs on a cluster with vastly more cores and
memory for bc, bfs, kcore, and sssp.
\revise{The main takeaway is that \emph{\xpointshort enables analytics on
massive graphs using shared-memory frameworks out-of-the-box while yielding
performance comparable or better than that of a cluster with the same resources
as the framework may support more efficient algorithms}.}

%% file: tbl_stampede_vs_optane.tex
\begin{table}[t]
\footnotesize
\centering
  \caption{Execution time (sec) of benchmarks in Galois on \xpointshort(OB)
  machine using efficient algorithms (non-vertex, asynchronous) and D-Galois on 
  Stampede cluster (DM) using vertex programs with minimum number of hosts
  that hold the graph (5 hosts for clueweb12, and uk14, and 20 hosts
  for wdc12). Speedup of \xpointshort over Stampede. 
  Best times highlighted in green.}
    
\label{tbl:stampvsoptane}
\begin{tabular}{@{ }l|@{\hskip1.8pt}l|@{\hskip5pt}r@{\hskip5pt}r@{\hskip5pt}r@{ }}
\toprule
\textbf{Graph}                     & \textbf{App}           & {\begin{tabular}[l]{@{}c@{}}\textbf{Stampede}\\\textbf{(DM)}\end{tabular}}   & {\begin{tabular}[l]{@{}c@{}}\textbf{\xpointshortend}\\\textbf{(OB)}\end{tabular}} & {\begin{tabular}[l]{@{}c@{}}\textbf{Speedup}\\\textbf{(DM/OB)}\end{tabular}}                    \\
\midrule                                                                                                       
\multirow{6}{*}{\textbf{clueweb12}}
                                &\textsf{bc}                 &\colorbox{white!20}{51.63}               & \colorbox{green!20}{12.68}                     & 4.07$\times$                           \\
                                &\textsf{bfs}                &\colorbox{white!20}{10.71}               & \colorbox{green!20}{6.43  }                    & 1.67$\times$                        \\
                                &\textsf{cc}                 &\colorbox{white!20}{13.70}               & \colorbox{green!20}{11.08}                     & 1.24$\times$                               \\
                                &\textsf{kcore}              &\colorbox{white!20}{186.03}              & \colorbox{green!20}{51.05}                     & 3.64$\times$                           \\
                                &\textsf{pr}                 &\colorbox{green!20}{155.00}              & \colorbox{white!20}{385.64}                    & 0.40$\times$                              \\
                                &\textsf{sssp}               &\colorbox{white!20}{33.87}               & \colorbox{green!20}{16.58}                     & 2.04$\times$                             \\

\midrule                                                                                                       
\multirow{6}{*}{\textbf{uk14}}
                                &\textsf{bc}                 & \colorbox{white!20}{172.23}            & \colorbox{green!20}{11.53}                      & 14.9$\times$                           \\
                                &\textsf{bfs}                & \colorbox{white!20}{28.38}             & \colorbox{green!20}{7.22}                       & 3.93$\times$                         \\
                                &\textsf{cc}                 & \colorbox{green!20}{14.56}             & \colorbox{white!20}{21.30}                      & 0.68$\times$                               \\
                                &\textsf{kcore}              & \colorbox{white!20}{56.08}             & \colorbox{green!20}{7.94}                       & 7.06$\times$                           \\
                                &\textsf{pr}                 & \colorbox{green!20}{82.77}             & \colorbox{white!20}{254.95}                     & 0.32$\times$                              \\
                                &\textsf{sssp}               & \colorbox{white!20}{52.49}             & \colorbox{green!20}{39.99 }                     & 1.31$\times$                             \\

\midrule                                                                                                       
\multirow{6}{*}{\revise{\textbf{iso\_m100}}}
                                &\textsf{bc}                 & \colorbox{green!20}{6.97}            & \colorbox{white!20}{11.57}                      & 0.60$\times$                           \\
                                &\textsf{bfs}                & \colorbox{white!20}{7.94}             & \colorbox{green!20}{3.69}                       & 2.15$\times$                         \\
                                &\textsf{cc}                 & \colorbox{green!20}{16.32}             & \colorbox{white!20}{23.69}                      & 0.69$\times$                               \\
                                &\textsf{kcore}              & \colorbox{white!20}{1.21}             & \colorbox{green!20}{0.48}                       &2.52$\times$                           \\
                                &\textsf{pr}                 & \colorbox{green!20}{191.21}             & \colorbox{white!20}{824.54}                     & 0.23$\times$                              \\
                                &\textsf{sssp}               & \colorbox{white!20}{61.90}             & \colorbox{green!20}{15.66}                     & 3.95$\times$                             \\

\midrule                                                                                                       
\multirow{6}{*}{\textbf{wdc12}}
                                &\textsf{bc}                 & \colorbox{white!20}{775.84}              &  \colorbox{green!20}{56.48}                   & 13.7$\times$                           \\
                                &\textsf{bfs}                & \colorbox{white!20}{71.50 }              &  \colorbox{green!20}{35.25}                   & 2.03$\times$                          \\
                                &\textsf{cc}                 & \colorbox{green!20}{69.21}               &  \colorbox{white!20}{76.00}                   & 0.91$\times$                               \\
                                &\textsf{kcore}              & \colorbox{white!20}{105.42}              &  \colorbox{green!20}{49.22}                   & 2.14$\times$                           \\
                                &\textsf{pr}                 & \colorbox{green!20}{118.01}              &  \colorbox{white!20}{1706.35}                 & 0.07$\times$                              \\
                                &\textsf{sssp}               & \colorbox{white!20}{136.47}              &  \colorbox{green!20}{118.81}                  & 1.15$\times$                              \\

\bottomrule
\end{tabular}
\end{table}

%% file: outofcore.tex
In addition to our main shared-memory experiments, we use an out-of-core graph
analytics system with app-direct mode on \xpointshort to \revise{determine if
an out-of-core system that uses \xpointshortend as external storage is
competitive with a shared-memory framework that uses it as main memory}.

\input{tbl_gridgraph_vs_optane}

\noindent{\textbf{Setup.}}
\revise{We use the state-of-the-art out-of-core graph
analytics framework, GridGraph~\cite{gridgraph} 
(some recent out-of-core
graph analytics frameworks~\cite{wonderland,lumos} 
are faster than GridGraph 
but they handle only a subset of algorithms that GridGraph handles).}
We compare GridGraph in \xpointshortend's app-direct (AD) with 
Galois in memory
mode (MM).
The \xpointshort machine was configured in AD mode as described in
Section~\ref{3dxpoint}. In AD, GridGraph manages the available DRAM
(memory budget given as 384GB) unlike in MM where DRAM is managed by the OS as
another cache level. The input graphs (preprocessed by GridGraph) are stored
on the \xpointshort modules which are used by GridGraph during execution.
We used a 512 by 512 grid as the partitioning grid for GridGraph (the GridGraph
paper used larger grid partitions for larger graphs to better fit blocks into
cache).~\footnote{We have tried larger grids, but preprocessing
fails as GridGraph opens more file descriptors than the machine supports.}
GridGraph uses a \texttt{signed 32-bit int} for storing the node IDs, making it
impractical for large graphs with $> 2^{31} - 1$ nodes such as wdc12. We
conduct a run of bfs and cc: it does not have bc, kcore, or sssp, and we have
observed pr failing due to assertion errors in the code.

\noindent{\textbf{Results.}}
Table~\ref{tbl:outofcore} compares the performance of \xpointshort in memory
mode (MM) running shared-memory Galois and app-direct mode (AD) running
out-of-core GridGraph. We observe that Galois using MM is orders of magnitude
faster than GridGraph (GridGraph bfs on uk14 failed to finish in 2 hours). This
can be attributed to the more sophisticated algorithms (in particular,
non-vertex programs and asynchronous data-driven algorithms supported
in Galois using MM unlike out-of-core frameworks that only
support vertex-programs) and the additional IO overhead required by
out-of-core frameworks, especially for real-world web-crawls with very high
diameter such as clueweb12 (diameter $\approx 500$).  We note that after few
rounds of computation on bfs for clueweb12, very few nodes get updated:
however, the blocks containing its corresponding edges still
must be loaded from the storage to be processed.
\revise{Note other out-of-core systems~\cite{lumos, wonderland} that 
are optimized for a subset of graph algorithms are
no more than an order of magnitude faster than GridGraph; therefore,
we expect Galois to outperform those systems in \xpointshort as well.}

\revise{To summarize, although out-of-core graph analytics systems can
also use \xpointshort via app-direct mode, the lack
of expressibility and the IO requirement of out-of-core systems
hurt runtime compared to a shared-memory framework that does not
have these limitations.}

%% file: tbl_gridgraph_vs_optane.tex
\begin{table}[t]
\footnotesize
\centering
  \caption{Execution time (sec) of benchmarks in Galois on \xpointshort in
  Memory Mode (MM) and the state-of-the-art out-of-core graph analytics
  framework GridGraph on \xpointshort in App-direct Mode (AD). A 512 by
  512 partition grid was used for GridGraph. Best times are highlighted in green. "---"
  indicates that system failed to finish in 2 hours.}

\label{tbl:outofcore}
\begin{tabular}{@{ }l|@{\hskip1.8pt}l|@{\hskip5pt}r@{\hskip5pt}r@{\hskip5pt}r@{ }}
\toprule
\textbf{Graph}                     & \textbf{App}           & {\begin{tabular}[l]{@{}c@{}}\textbf{GridGraph}\\\textbf{(AD)}\end{tabular}}   & {\begin{tabular}[l]{@{}c@{}}\textbf{Galois}\\\textbf{(MM)}\end{tabular}} & {\begin{tabular}[l]{@{}c@{}}\textbf{Speedup}\\\textbf{(AD/MM)}\end{tabular}}                    \\
\midrule
\multirow{2}{*}{\textbf{clueweb12}}
                                &\textsf{bfs}                &\colorbox{white!20}{5722.75}               & \colorbox{green!20}{6.43}                      & 890.0$\times$                        \\
                                &\textsf{cc}                 &\colorbox{white!0}{5411.23}               & \colorbox{green!20}{11.08}                     & 488.4$\times$                               \\
\midrule
\multirow{2}{*}{\textbf{uk14}}
                                &\textsf{bfs}                & \colorbox{white!20}{---}                   & \colorbox{green!20}{7.22}                       & NA                         \\
                                &\textsf{cc}                 & \colorbox{white!20}{5700.48}             & \colorbox{green!20}{21.30}                      & 267.6$\times$                               \\
\bottomrule
\end{tabular}
\end{table}

%% file: results_summary.tex
\revise{Our experiments show that graph analytics on \xpointshort is
competitive with analytics on both DRAM and distributed clusters. This is
important as the monetary cost of \xpointshort 
technology is much less than that of large DRAM or
machine clusters. In addition, unlike distributed execution which requires a
system designed for distributed analytics 
that typically restricts algorithmic choices 
for graph applications, 
\xpointshort allows running shared
memory graph analytics \emph{without changes to existing programs}. 

Our study only uses a selection of existing graph benchmarks that exist in different existing graph analytics systems, but
our findings generalize further to 
other graph analytics benchmarks, 
including those with more attributes similar to
edge weights in sssp. 
Additional attributes will increase the memory footprint
and require more memory accesses: therefore, our principles of reducing memory
accesses with efficient data-driven algorithms, accounting for near-memory
cache hit rates, and intelligent NUMA allocation would matter even more with
the increased memory footprint.}

While our study was specific to \xpointshortend, the guidelines in summarized
below \emph{apply to other large-memory analytics systems as well.} 

\begin{closeitemize}
\item Studies using synthetic power-law graphs like kron and rmat can be
misleading because unlike these graphs, large real-world web-crawls have large
diameters (Section~\ref{subsec:algo_design}).

\item For good performance on large diameter graphs, the programming model must
allow application developers to write work-efficient algorithms that need not
be vertex programs, and the system must provide data structures for sparse
worklists to enable asynchronous data-driven algorithms to be implemented
easily (Section~\ref{subsec:algo_design}).

\item On large-memory NUMA systems, the runtime must manage memory allocation
instead of delegating it to the OS. It must exploit huge pages and NUMA blocked
allocation. NUMA migration is not useful. (Section~\ref{subsec:3dxpoint_setup})

\end{closeitemize}

\revise{Finally, our study identifies several avenues 
for future work. 
As \xpointshort suffers if near-memory is not utilized well,
techniques can be developed to improve near-memory hit rate
to increase efficiency of graph analytics on \xpointshortend. 
In addition, this study focused on memory mode 
and showed graph analytics in memory mode is faster 
than that in app-direct mode. 
Work remains to be done to
determine the best manner of using app-direct mode for out-of-core graph analytics systems.}

%

%% file: related.tex
\noindent{\bf Shared-Memory Graph Processing.}
Shared-memory graph processing frameworks such as Galois~\cite{galois},
Ligra~\cite{ligra,gbbs}, 
and GraphIt~\cite{graphit} provide
users with abstractions to do graph computations that leverage 
a machine's underlying properties such as NUMA, memory locality, and
multicores. Shared-memory frameworks are limited by the 
available main memory on the system in which it loads the graph into memory
for processing: if a graph cannot fit, then out-of-core or distributed
processing must be used. However, if the graph fits in memory, the cost of shared
memory systems is less than out-of-core or distributed systems as they do
not suffer disk reading overhead or communication overhead, respectively.

\xpointshort increases the memory available to shared-memory graph
processing systems, and our evaluation shows that algorithms run with
\xpointshort are competitive or better than D-Galois~\cite{gluon}, a state-of-the-art
distributed graph analytics system. This is consistent with past work in which
it was shown that shared-memory graph processing on large graphs can be
efficient~\cite{gbbs}, and our findings extend to cases where a user has large
amounts of main memory (it is not limited to \xpointshortend).

\noindent{\bf Out-of-core Graph Processing.}
Out-of-core graph processing systems such as
GraphChi~\cite{graphchi}, X-Stream~\cite{xstream},
GridGraph~\cite{gridgraph}, Mosaic~\cite{mosaic}, Lumos~\cite{lumos}, CLIP~\cite{squeezeooc},
and BigSparse~\cite{bigsparse} compute by
loading appropriate portions of a graph into memory and writing back out
to disk in a disciplined manner to reduce disk overhead. Therefore,
these systems are not limited by main memory like shared-memory systems. The
overhead of disk operations, however, greatly impacts performance 
compared to shared-memory systems.


\noindent{\bf Distributed Graph Processing.}
Distributed graph processing systems such as
PowerGraph~\cite{powergraph}, Gemini~\cite{gemini}, D-Galois~\cite{gluon}, 
and others~\cite{giraph,combblas,powerlyra,pgxd,pregel,graphx} process large
graphs by distributing the graph among many machines which increases both
available memory and computational power. However,
communication among the machines is required,
and this can add significant runtime overhead.
Additionally, getting access to a distributed cluster can be expensive to
an average user.

\noindent{\bf Persistent Memory.}
Prior work on non-volatile memory includes file systems designed for
persistent memory~\cite{bpfs,nova,pmfs,hinfs}, making sure access to
persistent memory is efficient while being semantically
consistent~\cite{nvheaps,mnemosyne,nvmhptransaction,kaminotx}, database
systems in persistent
memory~\cite{writebehindlog,nvmreplicationdatastruct,dbnvm}.

\noindent{\bf \xpointshort Evaluation.} 
Many studies have evaluated the
potential of \xpointshort for different application domains.  Izraelevitz et
al.~\cite{3dxperf} presented a detailed analysis of the performance
characteristics of \xpointshort with evaluation on SPEC 2017 benchmarks,
various file systems, and databases.  \xpointshort has also been evaluated for
HPC applications with high memory and I/O bottlenecks~\cite{optaneSC19,
optaneHPC}. 
Malicevic et al.~\cite{nvmgraph} use app-direct-like mode to study graph 
analytics applications using emulated NVM system. However, they only 
study vertex programs on very small graphs.
Peng et al.~\cite{pengMemsys19} evaluates the performance of graph
analytics applications on \xpointshort in memory mode. However, they only use 
artificial Kronecker~\cite{kron} and RMAT~\cite{rmat} generated graphs, which, as we have
shown in this work, exhibit different structural properties compared to
real-world graphs.

Our study evaluates graph applications on large real-world graphs using 
efficient and sophisticated algorithms (in particular, non-vertex programs
and asynchronous data-driven algorithms) and is also
the first work, to our knowledge, to compare the performance of 
graph analytics applications on \xpointshort using Galois with 
the state-of-the-art distributed graph analytics framework,
D-Galois~\cite{gluon}, on a production level cluster (Stampede~\cite{stampede})
as well as to the out-of-core framework, GridGraph~\cite{gridgraph}, 
on \xpointshortend.

%% file: conclusion.tex
This paper proposed guidelines for high\hyp{}performance graph
analytics on Intel's \xpointshort memory and highlighted the principles that
apply to graph analytics for all large-memory settings.  In particular, our
study shows the importance of NUMA-aware memory allocation at the
application level and avoiding kernel overheads for \xpointshort as poor
applications of both concepts are more expensive on \xpointshort than on DRAM.
In addition, it shows the importance of non-vertex, asynchronous graph
algorithms for large memory systems as synchronous vertex programs do not scale
well as graphs grow.  We believe that \xpointshort is a viable
alternative to clusters with similar computational power for graph analytics
because they support a wider range of efficient algorithms while providing
competitive end-to-end performance.



%% file: ms.bbl
\begin{thebibliography}{10}

\bibitem{graph500}
{Graph 500 Benchmarks}, 2017.

\bibitem{squeezeooc}
Z.~Ai, M.~Zhang, Y.~Wu, X.~Qian, K.~Chen, and W.~Zheng.
\newblock {Squeezing out All the Value of Loaded Data: An Out-of-core Graph
  Processing System with Reduced Disk I/O}.
\newblock In {\em 2017 {USENIX} Annual Technical Conference ({USENIX} {ATC}
  17)}, pages 125--137, Santa Clara, CA, July 2017. {USENIX} Association.

\bibitem{writebehindlog}
J.~Arulraj, M.~Perron, and A.~Pavlo.
\newblock Write-behind logging.
\newblock {\em Proc. VLDB Endow.}, 10(4):337--348, Nov. 2016.

\bibitem{protein}
A.~Azad, G.~A. Pavlopoulos, C.~A. Ouzounis, N.~C. Kyrpides, and A.~Buluç.
\newblock {HipMCL: a high-performance parallel implementation of the Markov
  clustering algorithm for large-scale networks}.
\newblock {\em Nucleic Acids Research}, 46(6):e33--e33, 01 2018.

\bibitem{dobfs}
S.~Beamer, K.~Asanovi\'{c}, and D.~Patterson.
\newblock Direction-optimizing breadth-first search.
\newblock In {\em Proceedings of the International Conference on High
  Performance Computing, Networking, Storage and Analysis}, SC '12, pages
  12:1--12:10, Los Alamitos, CA, USA, 2012. IEEE Computer Society Press.

\bibitem{gap}
S.~Beamer, K.~Asanovic, and D.~A. Patterson.
\newblock The {GAP} benchmark suite.
\newblock {\em CoRR}, abs/1508.03619, 2015.

\bibitem{BRSLLP}
P.~Boldi, M.~Rosa, M.~Santini, and S.~Vigna.
\newblock Layered label propagation: A multiresolution coordinate-free ordering
  for compressing social networks.
\newblock In S.~Srinivasan, K.~Ramamritham, A.~Kumar, M.~P. Ravindra,
  E.~Bertino, and R.~Kumar, editors, {\em Proceedings of the 20th international
  conference on World Wide Web}, pages 587--596. ACM Press, 2011.

\bibitem{BoVWFI}
P.~Boldi and S.~Vigna.
\newblock The {W}eb{G}raph framework {I}: {C}ompression techniques.
\newblock In {\em Proc. of the Thirteenth International World Wide Web
  Conference (WWW 2004)}, pages 595--601, Manhattan, USA, 2004. ACM Press.

\bibitem{cvc2d}
E.~G. Boman, K.~D. Devine, and S.~Rajamanickam.
\newblock Scalable matrix computations on large scale-free graphs using 2d
  graph partitioning.
\newblock In {\em 2013 SC - International Conference for High Performance
  Computing, Networking, Storage and Analysis (SC)}, pages 1--12, Nov 2013.

\bibitem{combblas}
A.~Buluc and J.~R. Gilbert.
\newblock The combinatorial blas: Design, implementation, and applications.
\newblock {\em Int. J. High Perform. Comput. Appl.}, 25(4):496--509, Nov. 2011.

\bibitem{rmat}
D.~Chakrabarti, Y.~Zhan, and C.~Faloutsos.
\newblock {\em R-MAT: A Recursive Model for Graph Mining}, pages 442--446.

\bibitem{chaoticrelax}
D.~Chazan and W.~Miranker.
\newblock Chaotic relaxation.
\newblock {\em Linear Algebra and its Applications}, 2(2):199 -- 222, 1969.

\bibitem{powerlyra}
R.~Chen, J.~Shi, Y.~Chen, and H.~Chen.
\newblock Powerlyra: Differentiated graph computation and partitioning on
  skewed graphs.
\newblock In {\em Proceedings of the Tenth European Conference on Computer
  Systems}, EuroSys '15, pages 1:1--1:15, New York, NY, USA, 2015. ACM.

\bibitem{hinfs}
Y.~Chen, J.~Shu, J.~Ou, and Y.~Lu.
\newblock Hinfs: A persistent memory file system with both buffering and
  direct-access.
\newblock {\em ACM Trans. Storage}, 14(1):4:1--4:30, Apr. 2018.

\bibitem{nvheaps}
J.~Coburn, A.~M. Caulfield, A.~Akel, L.~M. Grupp, R.~K. Gupta, R.~Jhala, and
  S.~Swanson.
\newblock Nv-heaps: Making persistent objects fast and safe with
  next-generation, non-volatile memories.
\newblock In {\em Proceedings of the Sixteenth International Conference on
  Architectural Support for Programming Languages and Operating Systems},
  ASPLOS XVI, pages 105--118, New York, NY, USA, 2011. ACM.

\bibitem{bpfs}
J.~Condit, E.~B. Nightingale, C.~Frost, E.~Ipek, B.~Lee, D.~Burger, and
  D.~Coetzee.
\newblock Better i/o through byte-addressable, persistent memory.
\newblock In {\em Proceedings of the ACM SIGOPS 22Nd Symposium on Operating
  Systems Principles}, SOSP '09, pages 133--146, New York, NY, USA, 2009. ACM.

\bibitem{cormen01}
T.~Cormen, C.~Leiserson, R.~Rivest, and C.~Stein, editors.
\newblock {\em Introduction to Algorithms}.
\newblock {MIT} Press, 2001.

\bibitem{vpp}
I.~Corporation.
\newblock Intel{\textsuperscript{\textregistered}} vtune\textsuperscript{TM}
  platform profiler analysis, 2019.

\bibitem{vtune}
I.~Corporation.
\newblock Intel\textsuperscript{{\textregistered}} vtune\textsuperscript{TM}
  amplifier, 2019.

\bibitem{ipmctl}
I.~Corporation.
\newblock Ipmctl \xpointfullend, 2019.

\bibitem{kcore}
N.~S. {Dasari}, R.~{Desh}, and M.~{Zubair}.
\newblock Park: An efficient algorithm for k-core decomposition on multicore
  processors.
\newblock In {\em 2014 IEEE International Conference on Big Data (Big Data)},
  pages 9--16, Oct 2014.

\bibitem{gluon}
R.~Dathathri, G.~Gill, L.~Hoang, H.-V. Dang, A.~Brooks, N.~Dryden, M.~Snir, and
  K.~Pingali.
\newblock Gluon: A communication-optimizing substrate for distributed
  heterogeneous graph analytics.
\newblock In {\em Proceedings of the 39th ACM SIGPLAN Conference on Programming
  Language Design and Implementation}, PLDI '18, pages 752--768, New York, NY,
  USA, 2018. ACM.

\bibitem{gbbs}
L.~Dhulipala, G.~E. Blelloch, and J.~Shun.
\newblock Theoretically efficient parallel graph algorithms can be fast and
  scalable.
\newblock In {\em Proceedings of the 30th on Symposium on Parallelism in
  Algorithms and Architectures}, SPAA '18, pages 393--404, New York, NY, USA,
  2018. ACM.

\bibitem{pmfs}
S.~R. Dulloor, S.~Kumar, A.~Keshavamurthy, P.~Lantz, D.~Reddy, R.~Sankaran, and
  J.~Jackson.
\newblock System software for persistent memory.
\newblock In {\em Proceedings of the Ninth European Conference on Computer
  Systems}, EuroSys '14, pages 15:1--15:15, New York, NY, USA, 2014. ACM.

\bibitem{partitioningstudy}
G.~Gill, R.~Dathathri, L.~Hoang, and K.~Pingali.
\newblock {A Study of Partitioning Policies for Graph Analytics on Large-scale
  Distributed Platforms}.
\newblock volume~12 of {\em PVLDB}, 2018.

\bibitem{giraph}
Apache {G}iraph.
\newblock http://giraph.apache.org/, 2013.

\bibitem{powergraph}
J.~E. Gonzalez, Y.~Low, H.~Gu, D.~Bickson, and C.~Guestrin.
\newblock {PowerGraph: Distributed Graph-parallel Computation on Natural
  Graphs}.
\newblock In {\em Proceedings of the 10th USENIX Conference on Operating
  Systems Design and Implementation}, OSDI'12, pages 17--30, Berkeley, CA, USA,
  2012. USENIX Association.

\bibitem{disttc}
L.~Hoang, V.~Jatala, X.~Chen, U.~Agarwal, R.~Dathathri, G.~Gill, and
  K.~Pingali.
\newblock {DistTC: High Performance Distributed Triangle Counting}.
\newblock In {\em 2019 IEEE High Performance Extreme Computing Conference (HPEC
  2019)}, HPEC '19. IEEE, 2019.

\bibitem{mrbc}
L.~Hoang, M.~Pontecorvi, R.~Dathathri, G.~Gill, B.~You, K.~Pingali, and
  V.~Ramachandran.
\newblock {A Round-Efficient Distributed Betweenness Centrality Algorithm}.
\newblock In {\em Proceedings of the 24th ACM SIGPLAN Symposium on Principles
  and Practice of Parallel Programming (PPoPP'19)}, PPoPP, New York, NY, USA,
  2019. ACM.

\bibitem{pgxd}
S.~Hong, S.~Depner, T.~Manhardt, J.~Van Der~Lugt, M.~Verstraaten, and H.~Chafi.
\newblock Pgx.d: A fast distributed graph processing engine.
\newblock In {\em Proceedings of the International Conference for High
  Performance Computing, Networking, Storage and Analysis}, SC '15, pages
  58:1--58:12, New York, NY, USA, 2015. ACM.

\bibitem{3dxperf}
J.~Izraelevitz, J.~Yang, L.~Zhang, J.~Kim, X.~Liu, A.~Memaripour, Y.~J. Soh,
  Z.~Wang, Y.~Xu, S.~R. Dulloor, J.~Zhao, and S.~Swanson.
\newblock Basic performance measurements of the intel optane {DC} persistent
  memory module.
\newblock {\em CoRR}, abs/1903.05714, 2019.

\bibitem{bigsparse}
S.~W. Jun, A.~Wright, S.~Zhang, S.~Xu, and Arvind.
\newblock Bigsparse: High-performance external graph analytics.
\newblock {\em CoRR}, abs/1710.07736, 2017.

\bibitem{nvmhptransaction}
A.~Kolli, S.~Pelley, A.~Saidi, P.~M. Chen, and T.~F. Wenisch.
\newblock High-performance transactions for persistent memories.
\newblock In {\em Proceedings of the Twenty-First International Conference on
  Architectural Support for Programming Languages and Operating Systems},
  ASPLOS '16, pages 399--411, New York, NY, USA, 2016. ACM.

\bibitem{graphchi}
A.~Kyrola, G.~Blelloch, and C.~Guestrin.
\newblock Graphchi: Large-scale graph computation on just a pc.
\newblock In {\em Proceedings of the 10th USENIX Conference on Operating
  Systems Design and Implementation}, OSDI'12, pages 31--46, Berkeley, CA, USA,
  2012. USENIX Association.

\bibitem{kron}
J.~Leskovec, D.~Chakrabarti, J.~Kleinberg, C.~Faloutsos, and Z.~Ghahramani.
\newblock Kronecker graphs: An approach to modeling networks.
\newblock {\em J. Mach. Learn. Res.}, 11:985--1042, Mar. 2010.

\bibitem{mmap}
Linux.
\newblock Mmap.

\bibitem{numactl}
Linux.
\newblock Numactl.

\bibitem{thp}
Linux.
\newblock Transparent huge pages: Linux kernel.

\bibitem{mosaic}
S.~Maass, C.~Min, S.~Kashyap, W.~Kang, M.~Kumar, and T.~Kim.
\newblock Mosaic: Processing a trillion-edge graph on a single machine.
\newblock In {\em Proceedings of the Twelfth European Conference on Computer
  Systems}, EuroSys '17, pages 527--543, New York, NY, USA, 2017. ACM.

\bibitem{pregel}
G.~Malewicz, M.~H. Austern, A.~J. Bik, J.~C. Dehnert, I.~Horn, N.~Leiser, and
  G.~Czajkowski.
\newblock Pregel: a system for large-scale graph processing.
\newblock In {\em Proceedings ACM SIGMOD Intl Conf. on Management of Data},
  SIGMOD '10, pages 135--146, 2010.

\bibitem{nvmgraph}
J.~Malicevic, S.~Dulloor, N.~Sundaram, N.~Satish, J.~Jackson, and
  W.~Zwaenepoel.
\newblock {Exploiting NVM in Large-scale Graph Analytics}.
\newblock In {\em Proceedings of the 3rd Workshop on Interactions of NVM/FLASH
  with Operating Systems and Workloads}, INFLOW '15, pages 2:1--2:9, New York,
  NY, USA, 2015. ACM.

\bibitem{kaminotx}
A.~Memaripour, A.~Badam, A.~Phanishayee, Y.~Zhou, R.~Alagappan, K.~Strauss, and
  S.~Swanson.
\newblock Atomic in-place updates for non-volatile main memories with
  kamino-tx.
\newblock In {\em Proceedings of the Twelfth European Conference on Computer
  Systems}, EuroSys '17, pages 499--512, New York, NY, USA, 2017. ACM.

\bibitem{wdc}
R.~Meusel, S.~Vigna, O.~Lehmberg, and C.~Bizer.
\newblock Web data commons - hyperlink graphs, 2012.

\bibitem{meyer98}
U.~Meyer and P.~Sanders.
\newblock Delta-stepping: A parallel single source shortest path algorithm.
\newblock In {\em Proc. European Symposium on Algorithms}, ESA '98, pages
  393--404, 1998.

\bibitem{galois}
D.~Nguyen, A.~Lenharth, and K.~Pingali.
\newblock A lightweight infrastructure for graph analytics.
\newblock In {\em Proceedings of the Twenty-Fourth ACM Symposium on Operating
  Systems Principles}, SOSP '13, pages 456--471, New York, NY, USA, 2013. ACM.

\bibitem{pagerank}
L.~Page, S.~Brin, R.~Motwani, and T.~Winograd.
\newblock The pagerank citation ranking: Bringing order to the web.
\newblock Technical Report 1999-66, Stanford InfoLab, November 1999.
\newblock Previous number = SIDL-WP-1999-0120.

\bibitem{pengMemsys19}
I.~B. Peng, M.~B. Gokhale, and E.~W. Green.
\newblock System evaluation of the intel optane byte-addressable nvm.
\newblock In {\em Proceedings of the International Symposium on Memory
  Systems}, MEMSYS '19, pages 304--315, New York, NY, USA, 2019. ACM.

\bibitem{pingali11}
K.~Pingali, D.~Nguyen, M.~Kulkarni, M.~Burtscher, M.~A. Hassaan, R.~Kaleem,
  T.-H. Lee, A.~Lenharth, R.~Manevich, M.~M{\'e}ndez-Lojo, D.~Prountzos, and
  X.~Sui.
\newblock The {TAO} of parallelism in algorithms.
\newblock In {\em Proc. ACM SIGPLAN Conf. Programming Language Design and
  Implementation}, PLDI '11, pages 12--25, 2011.

\bibitem{clueweb}
T.~L. Project.
\newblock {The ClueWeb12 Dataset}, 2013.

\bibitem{xstream}
A.~Roy, I.~Mihailovic, and W.~Zwaenepoel.
\newblock X-stream: Edge-centric graph processing using streaming partitions.
\newblock In {\em Proceedings of the Twenty-Fourth ACM Symposium on Operating
  Systems Principles}, SOSP '13, pages 472--488, New York, NY, USA, 2013. ACM.

\bibitem{lpcc}
L.~G. Shapiro.
\newblock Connected component labeling and adjacency graph construction.
\newblock In T.~Y. Kong and A.~Rosenfeld, editors, {\em Topological Algorithms
  for Digital Image Processing}, volume~19 of {\em Machine Intelligence and
  Pattern Recognition}, pages 1 -- 30. North-Holland, 1996.

\bibitem{pjcc}
Y.~Shiloach and U.~Vishkin.
\newblock An o(logn) parallel connectivity algorithm.
\newblock {\em Journal of Algorithms}, 3(1):57 -- 67, 1982.

\bibitem{ligra}
J.~Shun and G.~E. Blelloch.
\newblock Ligra: a lightweight graph processing framework for shared memory.
\newblock In {\em Proceedings ACM SIGPLAN Symp. Principles and Practice of
  Parallel Programming}, PPoPP '13, pages 135--146, 2013.

\bibitem{stampede}
D.~Stanzione, B.~Barth, N.~Gaffney, K.~Gaither, C.~Hempel, T.~Minyard,
  S.~Mehringer, E.~Wernert, H.~Tufo, D.~Panda, and P.~Teller.
\newblock {Stampede 2: The Evolution of an XSEDE Supercomputer}.
\newblock In {\em Proceedings of the Practice and Experience in Advanced
  Research Computing 2017 on Sustainability, Success and Impact}, PEARC17,
  pages 15:1--15:8, New York, NY, USA, 2017. ACM.

\bibitem{shortcutcc}
S.~Stergiou, D.~Rughwani, and K.~Tsioutsiouliklis.
\newblock Shortcutting label propagation for distributed connected components.
\newblock In {\em Proceedings of the Eleventh ACM International Conference on
  Web Search and Data Mining}, WSDM '18, pages 540--546, New York, NY, USA,
  2018. ACM.

\bibitem{dbnvm}
A.~van Renen, V.~Leis, A.~Kemper, T.~Neumann, T.~Hashida, K.~Oe, Y.~Doi,
  L.~Harada, and M.~Sato.
\newblock Managing non-volatile memory in database systems.
\newblock In {\em Proceedings of the 2018 International Conference on
  Management of Data}, SIGMOD '18, pages 1541--1555, New York, NY, USA, 2018.
  ACM.

\bibitem{mnemosyne}
H.~Volos, A.~J. Tack, and M.~M. Swift.
\newblock Mnemosyne: Lightweight persistent memory.
\newblock In {\em Proceedings of the Sixteenth International Conference on
  Architectural Support for Programming Languages and Operating Systems},
  ASPLOS XVI, pages 91--104, New York, NY, USA, 2011. ACM.

\bibitem{lumos}
K.~Vora.
\newblock {{LUMOS}: Dependency-Driven Disk-based Graph Processing}.
\newblock In {\em 2019 {USENIX} Annual Technical Conference ({USENIX} {ATC}
  19)}, pages 429--442, Renton, WA, July 2019. {USENIX} Association.

\bibitem{optaneSC19}
M.~Weiland, H.~Brunst, T.~Quintino, N.~Johnson, O.~Iffrig, S.~Smart, C.~Herold,
  A.~Bonanni, A.~Jackson, and M.~Parsons.
\newblock An early evaluation of intel's optane dc persistent memory module and
  its impact on high-performance scientific applications.
\newblock In {\em Proceedings of the International Conference for High
  Performance Computing, Networking, Storage and Analysis}, SC '19, pages
  76:1--76:19, New York, NY, USA, 2019. ACM.

\bibitem{optaneHPC}
K.~Wu, F.~Ober, S.~Hamlin, and D.~Li.
\newblock Early evaluation of intel optane non-volatile memory with {HPC} {I/O}
  workloads.
\newblock {\em CoRR}, abs/1708.02199, 2017.

\bibitem{graphx}
R.~S. Xin, J.~E. Gonzalez, M.~J. Franklin, and I.~Stoica.
\newblock Graphx: A resilient distributed graph system on spark.
\newblock In {\em First International Workshop on Graph Data Management
  Experiences and Systems}, GRADES '13, 2013.

\bibitem{nova}
J.~Xu and S.~Swanson.
\newblock {NOVA}: {A Log-structured File System for Hybrid
  Volatile/Non-volatile Main Memories}.
\newblock In {\em 14th {USENIX} Conference on File and Storage Technologies
  ({FAST} 16)}, pages 323--338, Santa Clara, CA, 2016. {USENIX} Association.

\bibitem{nvmreplicationdatastruct}
M.~Zarubin, T.~Kissinger, D.~Habich, and W.~Lehner.
\newblock Efficient compute node-local replication mechanisms for nvram-centric
  data structures.
\newblock In {\em Proceedings of the 14th International Workshop on Data
  Management on New Hardware}, DAMON '18, pages 7:1--7:9, New York, NY, USA,
  2018. ACM.

\bibitem{wonderland}
M.~Zhang, Y.~Wu, Y.~Zhuo, X.~Qian, C.~Huan, and K.~Chen.
\newblock Wonderland: A novel abstraction-based out-of-core graph processing
  system.
\newblock In {\em Proceedings of the Twenty-Third International Conference on
  Architectural Support for Programming Languages and Operating Systems},
  ASPLOS ’18, page 608–621, New York, NY, USA, 2018. Association for
  Computing Machinery.

\bibitem{graphit}
Y.~Zhang, M.~Yang, R.~Baghdadi, S.~Kamil, J.~Shun, and S.~Amarasinghe.
\newblock Graphit: A high-performance graph dsl.
\newblock {\em Proc. ACM Program. Lang.}, 2(OOPSLA):121:1--121:30, Oct. 2018.

\bibitem{gemini}
X.~Zhu, W.~Chen, W.~Zheng, and X.~Ma.
\newblock Gemini: {A Computation-centric Distributed Graph Processing System}.
\newblock In {\em Proceedings of the 12th USENIX Conference on Operating
  Systems Design and Implementation}, OSDI'16, pages 301--316, Berkeley, CA,
  USA, 2016. USENIX Association.

\bibitem{gridgraph}
X.~Zhu, W.~Han, and W.~Chen.
\newblock Gridgraph: Large-scale graph processing on a single machine using
  2-level hierarchical partitioning.
\newblock In {\em 2015 {USENIX} Annual Technical Conference ({USENIX} {ATC}
  15)}, pages 375--386, Santa Clara, CA, 2015. {USENIX} Association.

\end{thebibliography}
